\documentclass[3p]{elsarticle}
\usepackage{lineno}
\usepackage{epstopdf}
\usepackage{hyperref}
\usepackage{amsmath}
\usepackage{float}
\usepackage{amssymb}
\usepackage{subcaption}
\usepackage{graphicx}
\usepackage [english]{babel}
\usepackage [autostyle, english = american]{csquotes}
\MakeOuterQuote{"}

\begin{document}

\begin{frontmatter}

\title{Increasing charge lifetime in DC polarized electron guns by offsetting the anode}
\author[bnl]{Omer Rahman}
\author[bnl]{Erdong Wang\corref{cor2}}
\ead{wange@bnl.gov}
\author[bnl,sbu]{Ilan Ben-Zvi}
\author[sbu]{Jyoti Biswas}
\author[bnl]{John Skaritka}

\cortext[cor2]{Corresponding author}
\address[bnl]{Collider-Accelerator Department, Brookhaven National Lab, Building 911, P.O. Box 5000, Upton, NY-11973, USA.}
\address[sbu]{Department of Physics and Astronomy, Stony Brook University, Stony Brook, NY-11794, USA.}
\begin{abstract}
Charge lifetime of strained superlattice GaAs photocathodes in DC guns is limited by ion back bombardment. It needs to be improved at least an order of magnitude to meet the requirements for future colliders such as Electron-Ion Collider (EIC). In this work, we propose and present simulation results for an offset anode scheme to increase charge lifetime in DC guns. This scheme eliminates the bombardment of high energy ions on the cathode and enables maximum usage of the available cathode area. Depending on the size of the available cathode area, this method can increase the charge lifetime by an order of magnitude compared to the current best alternative method. An anode assembly capable of in-vacuum movement is required for this method, which has been designed and fabricated at Brookhaven National Laboratory.
\end{abstract}
\end{frontmatter}

\section{Introduction}
Strained Superlattice GaAs photocathode based high voltage DC (HV-DC) guns have been used as sources of polarized electron beam in various facilities around the world \cite{alley1995stanford,cates1989bates,hartmann1990source,jlab2007s}. These photocathodes can provide beam polarization up to 92\%, with quantum efficiency (QE) of about 1\% at 780 nm laser illumination \cite{nishitani2005highly}. The photocathodes need to be coated with Cs and an oxidant (O$_2$ or NF$_3$) to achieve "Negative Electron Affinity (NEA)" on the surface. This NEA layer is highly sensitive to the vacuum levels of the gun and should be operated at extremely high vacuum, possibly at pressures below 10$^{-11}$ Torr \cite{sinclair1999recent}. The degradation rate of this layer determines how much charge can be extracted from a photocathode before it becomes unusable. The two widely used metrics to characterize HV-DC polarized guns are the fluence lifetime and the charge lifetime. Fluence lifetime is defined as the amount of charge extracted per unit area from a cathode before the QE drops to 1/e of its initial value \cite{sinclair2001}. Under constant operating conditions, fluence lifetime is a constant for a photogun and can be used to compare the performances of different guns \cite{jlab2011}. Charge lifetime is defined as the total charge extracted from a photocathode until the quantum efficiency falls to 1/e of its initial value and can be used calculate the operation time of a gun \cite{aulenbacher2011polarized}. If the operating conditions are fixed, the fluence lifetime of a gun can be used to estimate the charge lifetime for different laser sizes. 
\par
For HV-DC polarized electron guns, ion back bombardment (IBB) is the dominant lifetime-limiting process for cathodes during operation \cite{aulenbacher1997mami}. In this process, the electron beam ionizes the residual gas as it propagates through DC gap and downstream beam line. The ions that are generated downstream, past the DC gap, can be blocked using a biased anode \cite{grames2008biased}. However, the ions generated in the DC gap cannot be blocked. These ions get accelerated towards the photocathode due to the DC electric field, eventually hitting the cathode surface. 
Operational experiences at several polarized electron beam facilities have shown clear damage pattern corresponding to bombardment from ions \cite{grames2005ion}. The SLC gun has demonstrated extraction of 10-16 nC bunch charge at $\mu$A levels of average current \cite{clendenin2003slac}. Jlab group has demonstrated 4 mA average current operation at 1500 MHz rep rate (pC bunch charge) with 85 C of charge lifetime \cite{adderley2011cebaf}. High bunch charge (nC level) with high average current (mA level) operation, with acceptable kC charge lifetime, is yet to be demonstrated and has been one of the main focus of research in the field of polarized electron guns.
For future colliders, such as an Electron-Ion Collider (EIC), the requirements for the beam parameters for polarized electron sources are beyond what has been achieved by any operational facility \cite{ptitsyn2017erl}\cite{bruening2013large}. The charge lifetime for a polarized gun also needs to be improved an order of magnitude for a realistic polarized electron source for the proposed cost effective ERL based EIC \cite{ptitsyn2016erl}. A high current polarized electron source can also be used in producing a high intensity polarized positron beam, which can be used in various collider applications \cite{abbott2016production}.\par
Apart from improving the vacuum levels in a gun, multiple operational techniques have been explored to increase the charge lifetime of a polarized gun. These techniques include increasing the laser spot size, operating the laser off center, limiting the cathode active area and biasing the anode to block ions coming from downstream of the DC gap \cite{grames2008biased}\cite{aulenbacher2011polarized}\cite{tsentalovich2014status}. 
Offsetting the laser for higher charge lifetime was discovered by Jefferson lab, which is currently the standard mode of operation for polarized electron guns \cite{jlab2007s}. In this scheme, beam is extracted from a spot radially offset from the electrostatic center (EC) of the cathode. This ensures that the QE degradation under the laser spot is only due to low energy ion bombardment, since the higher energy ions gets focused at the EC. The damage rate due to low energy ions are much slower compared to the damage due to higher energy ions and therefore this operation mode yields a higher charge lifetime \cite{jlab2011}. One of the major limitations for this scheme is that it limits the maximum laser spot size to be much less than 50\% of the available cathode area. Therefore, an opportunity of getting even higher charge lifetime by using a larger laser spot becomes impractical using offset laser scheme.  \par 
In this work, we propose an offset anode scheme to completely eliminate high energy ion bombardment on the cathode surface and make maximum use of the available photocathode area. In this scheme, the axial symmetry of cathode-anode configuration is broken by transversely offsetting the anode, while keeping the laser spot on the center of the cathode. Depending on the size of the available photocathode area, the charge lifetime can be increased a factor of ten as compared to the best alternate method.\par
The rest of this paper is arranged in three sections. The first section describes the detail of the offset anode scheme, compares it with other schemes and provides a brief description of a gun that is being designed to test the scheme. The second section focuses on the ion back bombardment simulation for various offset anode and offset laser operation. Finally, the third section discusses the beam dynamics simulation results for the offset anode scheme.
\section{Offset anode scheme}
\subsection{Description of the scheme}
In this section, we discuss the physics of ion back bombardment for three different schemes: on axis operation (both laser and anode on axis), offset laser (O.L.) operation and offset anode (O.A.) operation. The energy range of the back bombarded ions play a crucial role in determining the damage rate for different operation schemes. For clarity, we will use the following convention to group back bombarded ions based on their energy: low energy (0-30\% of maximum energy), medium energy (30\%-80\% of maximum energy) and high energy (80\%-100\% of maximum energy). It should be noted that the damage on the photocathode does not have clear thresholds in terms of ion energy and this convention is adopted for this particular paper to clarify the concepts and compare different operational schemes.  \par
When the laser is operated at the center of the cathode, the entire laser illuminated area gets bombarded by ions with energy varying from low to high energy. The high energy ions at 80\% to 100\% of the maximum energy, are focused at the EC due to the Pierce like focusing in the DC gap. The medium energy ions cannot be focused as much as the high energy ions since they are generated closer to the cathode and therefore have less distance to travel in the focusing channel. The low energy ions, ions that are generated within the first few mm from the cathode, are distributed over the entire laser illumination spot. Therefore, the entire laser illumination area is damaged to various degree depending on the energy of the bombarded ions. This mode of operation has poor operating conditions and shown to have poor fluence (charge) lifetime \cite{jlab2011}.\par

During offset laser operation, the area under laser illumination avoids the high energy ion bombardment. It has been shown experimentally that the larger the offset, the higher the charge lifetime \cite{jlab2011}. This experimental finding suggests that even though the laser spot is not bombarded by the high energy ions, the medium and low energy ions are detrimental enough towards degradation of cathode lifetime. As the offset amount increases, the laser illuminated area encounters less medium energy ions. This is equivalent of improving the operating conditions which yields a higher fluence (charge) lifetime. However, for the highest fluence lifetime in this scheme, the laser spot size gets limited due to two constraints: offset from EC and distance from the edge. The spot size has to be small enough such that it can fit on either side of the electrostatic center of the cathode and simultaneously be sufficiently far away from the edge of the photocathode. Beam extraction from too close to the edge may result in extreme trajectory of the beam which may cause beam loss at the anode or nearby beam pipe. This will create unwanted outgassing and QE will degrade much faster than on center operation \cite{jlab2010}. These constraints restrict the maximum size of the laser spot to be less than 50\% of the total available area of the photocathode. Grames et al. (2011) has carried out extensive experimental studies to understand the dependency of charge lifetime on laser spot size and laser offset amount \cite{jlab2011}. In their experiments, a 3.5 kC charge lifetime with 4 mA average current was achieved from a 1.5 mm laser spot, with 2 mm radial offset, from bulk GaAs at 532 nm laser operation. The total cathode area for this particular gun is 12.8 mm in diameter. Therefore a big part of the photocathode remains unused. The laser spot can be moved from spot to spot on the photocathode, but that will require stopping the gun operation, possible reactivation of the cathode using a mask on the desired spot and finally tuning the beam optics for the new laser position.  \par


\begin{figure}[t!]
\begin{minipage}[b]{.5\linewidth}
\centering
\includegraphics[width =0.9\textwidth]{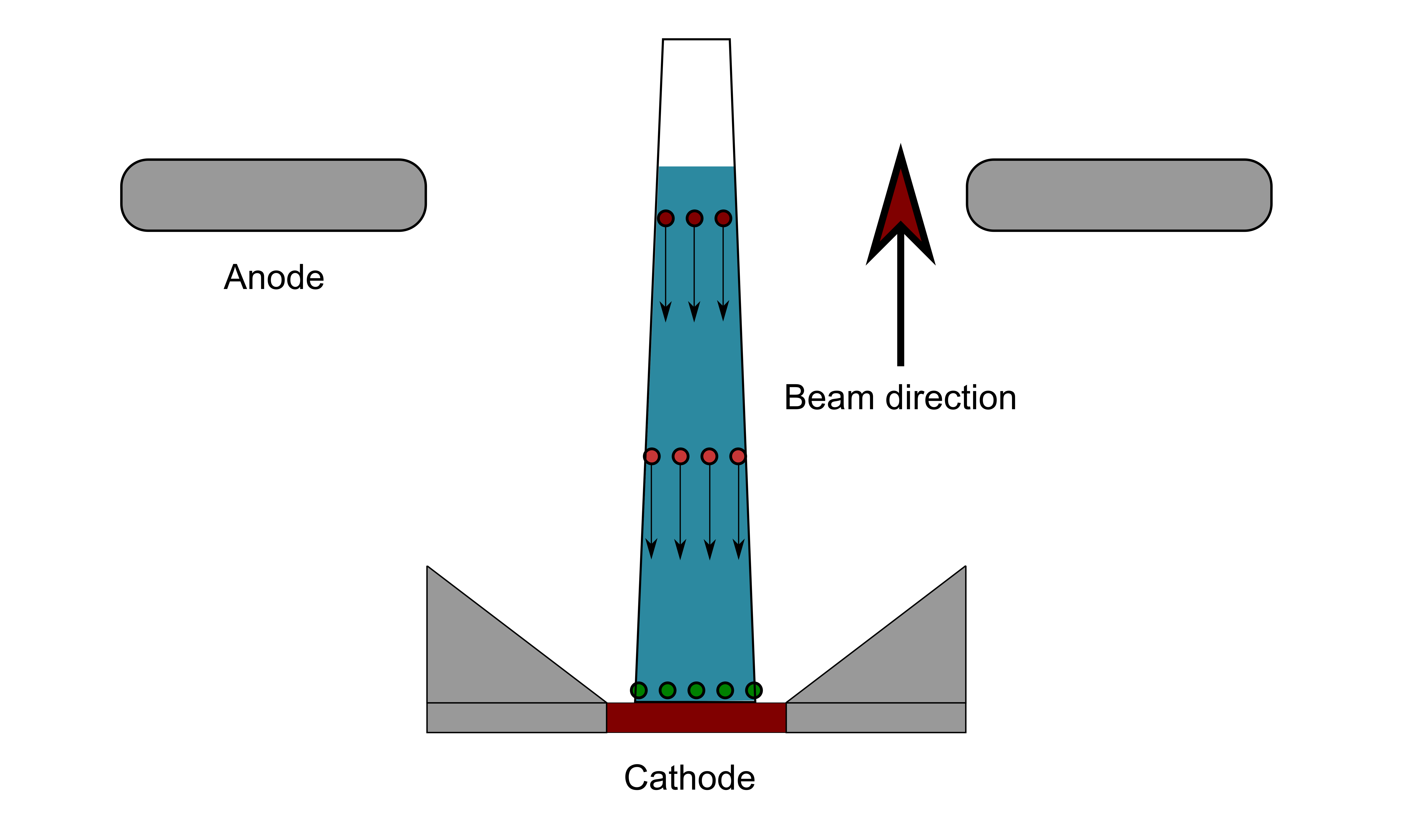}
\subcaption{On axis operation}\label{fig:1a}
\end{minipage}
\begin{minipage}[b]{.5\linewidth}
\centering
\includegraphics[width =\textwidth]{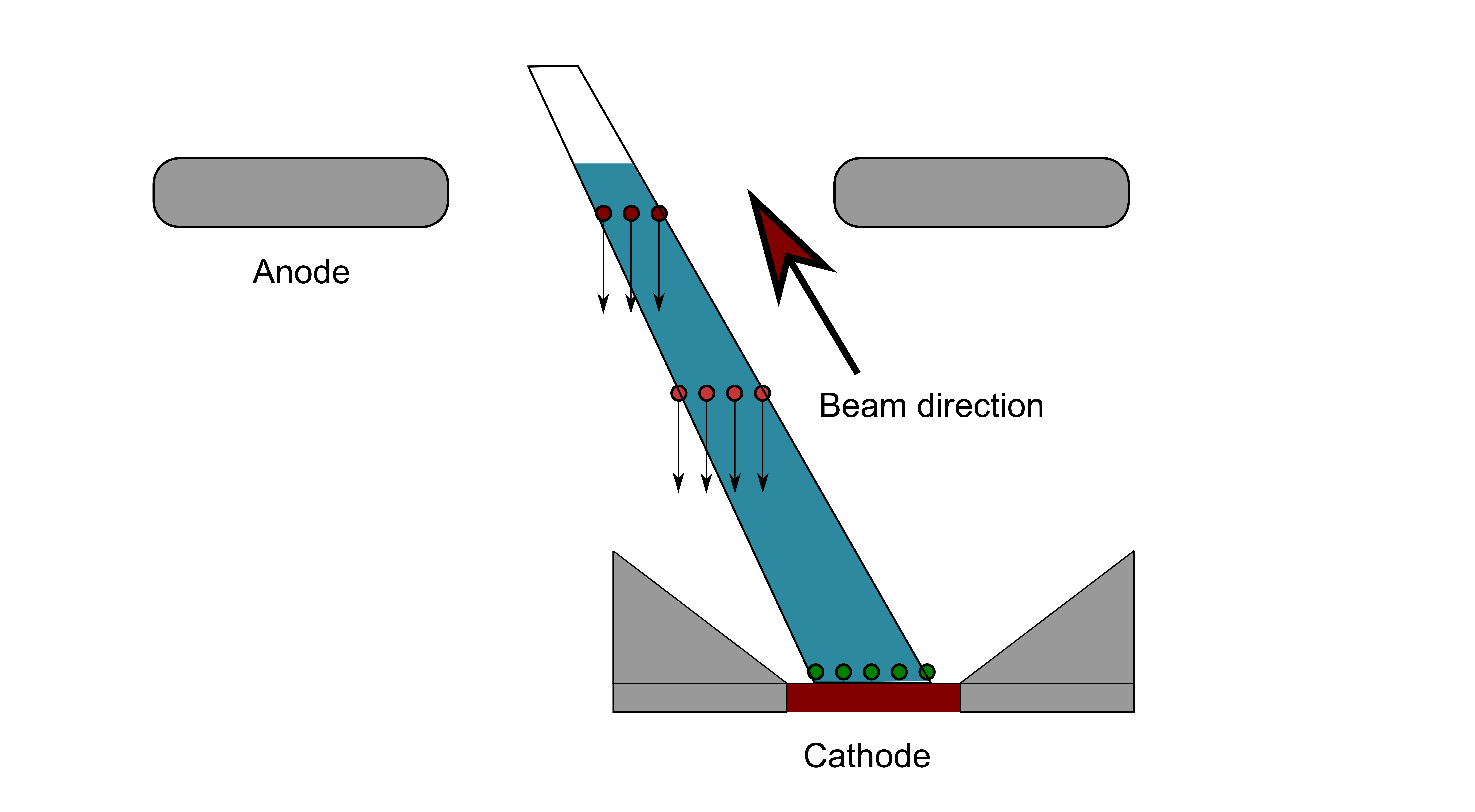}
\subcaption{Offset anode operation}\label{fig:1b}
\end{minipage}
\caption{Ion generation at different points along the beam path in the DC gap for on axis and offset anode operations. Green, light red and dark red circles indicate low, medium and high energy ions. The arrow on the ion circles indicates general direction towards which the ions tend to move. Effect of focusing on the medium and high energy ions are not shown. }
\label{fig:ion_gen}
\end{figure}

To maximize the use of available photocathode area, while maintaining the highest fluence lifetime, we propose to break the axial symmetry of the cathode-anode configuration by offsetting the anode transversely. Unlike offset laser scheme, we propose to operate the laser on the center of the cathode. A transverse magnetic field, generated by a Helmholtz coil, can deflect the electron beam to the center of the offset anode at the exit of the gun. After that, a set corrector magnets can bring the beam back to the center of the beam pipe.
\begin{figure}[h!]
 \centering   
  \includegraphics[width=0.9\textwidth]{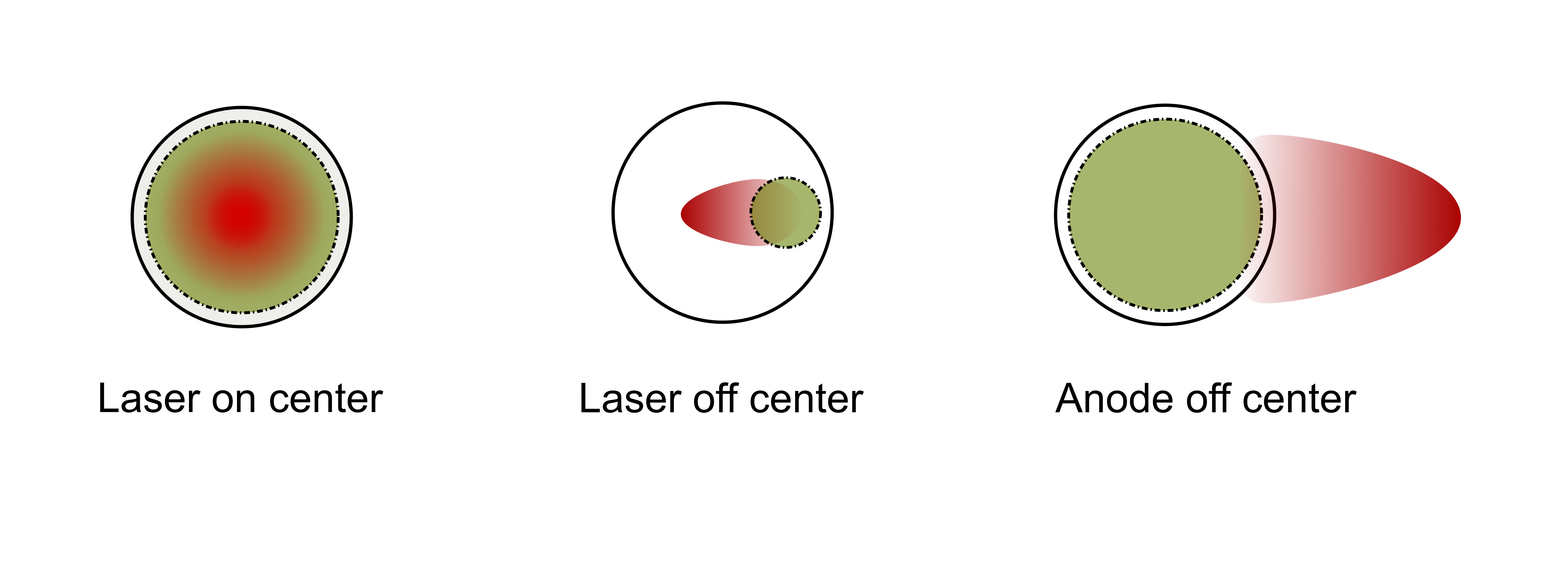}
   \caption{Laser spot size and expected damage profiles for various operation schemes. The green area indicates damage from low energy ions and the red area indicates damage from medium - high energy ions. Increasing intensity of the red section represents increasing energy. The solid and dashed circles respectively indicate total cathode area and laser spot size.}
    \label{fig:offset_scheme}
\end{figure}
The ions generated along the beam path will have different bombardment spots depending on the point of generation along the beam trajectory. The trajectory of the lower energy ions, generated close to the cathodes, will be similar compared to that of offset laser scheme and they will bombard the area under the laser spot. However the medium and high energy ions, generated further down the beam path, will have different trajectories compared to that of the other schemes. These ions will be generated at location transversely offset from the cathode and will completely miss the cathode given a sufficiently large offset. Figure \ref{fig:ion_gen} shows the possible ion generation at different points along the beam path for on axis and offset anode scheme. The transverse magnetic field in the DC gap will have minimal effect on the heavy Hydrogen ions. These ions should follow a nearly straight trajectory from the point of creation to the cathode surface. So if the anode is offset transversely by 6 mm, we should expect the high energy ions to hit the cathode area at about 6 mm offset from the center. The anode could be offset to a suitable distance such that the medium and high energy ions completely miss the photocathode surface area. Figure \ref{fig:offset_scheme} provides a visual representation of the laser spot position and energy profile of the back bombarded ions for on axis, offset laser and offset anode scheme.
Since the electrostatic center is now protected from medium-high energy ion bombardment, the cathode degradation will only be from low energy ions, similar to that of offset laser scheme. This is the exact operating condition which results in the highest fluence lifetime for offset laser scheme. However now that the laser spot can be operated at the electrostatic center, the laser spot size can be increased to maximize the usage of photocathode area. If ion back bombardment is the dominant mechanism of cathode degradation and there is no significant beam loss, the charge lifetime of a gun is directly proportional to the laser spot area. Therefore, this scheme will help increase the charge lifetime of a polarized electron gun.\par 

\begin{table}[h!]
\centering
\begin{tabular}{||c | c | c | c ||} 
 \hline
 Cathode radius & Max. laser spot (O.L.) & Max. laser spot (O.A.)  & Charge lifetime increment\\ [0.5ex]
 \hline
 mm & mm$^2$ & mm$^2$ & factor of \\
 \hline\hline
 8 & 7.065 & 78.5 & 11.11 \\ 
 \hline
 9 & 12.56 & 113.04 & 9 \\
 \hline
 10 & 19.625 & 153.86 & 7.84\\
 \hline
 11 & 28.26 & 200.96 & 7.11 \\ [1ex] 
 \hline
 12 & 38.465 & 254.34 & 6.61 \\ [1ex]
 \hline
 13 & 50.24 & 314 & 6.25 \\ [1ex]
 \hline\hline
\end{tabular}
\caption{Comparison of maximum laser spot sizes for offset laser and offset anode operation and expected charge lifetime increment factor. }
\label{table:beamdia}
\end{table}
If $r$ is the radius of the available cathode, $a$ is the offset from the EC and $b$ is the clearance from the edge of the cathode, the maximum available area for laser spot in offset laser operation will be $\pi(r-a-b)^2 /4$. For offset anode operation, the available area for laser illumination will be $\pi (r-b)^2$.  Experiments have shown that a 2-4 mm offset from the center and 3-4 mm clearance from the edge of the cathode maximizes the charge lifetime \cite{dunham2013record}\cite{jlab2011}. We can use these experimentally obtained parameters to estimate the increment of charge lifetime using offset anode scheme. The ratio of the areas can be used to estimate the increment of charge lifetime for offset anode scheme. As a conservative estimate, we use $a$ to be 2 mm and $b$ to be 3 mm. Table 1 shows the estimated charge lifetime increment for offset anode scheme compared to offset laser scheme.\par

We note that beam transport, including optimized design of cathode-anode assembly, is of prime importance for this scheme to be successful. However, the details of beam transport design is beyond the scope of this paper. In this particular paper, we will focus on proving the physics principle of the offset anode operation and its scope towards improving charge lifetime.
In order to prove the feasibility of this scheme, we simulated and studied ion back bombardment in a DC gun for various laser and anode offset positions. The simulation results show that medium-high energy ions could indeed be shifted in an offset anode scheme. The required offset will depend on the total cathode area of the gun. The beam transport and emittance dilution for this scheme, for this particular gun, were also studied. A 5.2 nC bunch with 8.6 mm laser spot diameter could be transported out of the gun with a 13 mm anode offset without much degradation in beam quality.  

\subsection{Description of the gun and simulation setup}
We used the BNL large cathode prototype gun and associated beam line for this simulation study \cite{wang2018high}. This gun is designed for a 26 mm diameter cathode, with a nominal DC gap of 5.6 cm. The maximum design voltage of this gun is 350 KV, providing 4 MV/m field on the center of the cathode. The aperture of the anode is 36 mm in diameter and the diameter of the beam line is 10 cm. The anode section in the DC gun vessel has been modified to accommodate a recessed structure that enables mechanical movements of the anode using a special actuator assembly.  \par
\begin{figure}[h!]
\begin{minipage}[b]{.5\linewidth}
\centering
\includegraphics[width =0.75\textwidth]{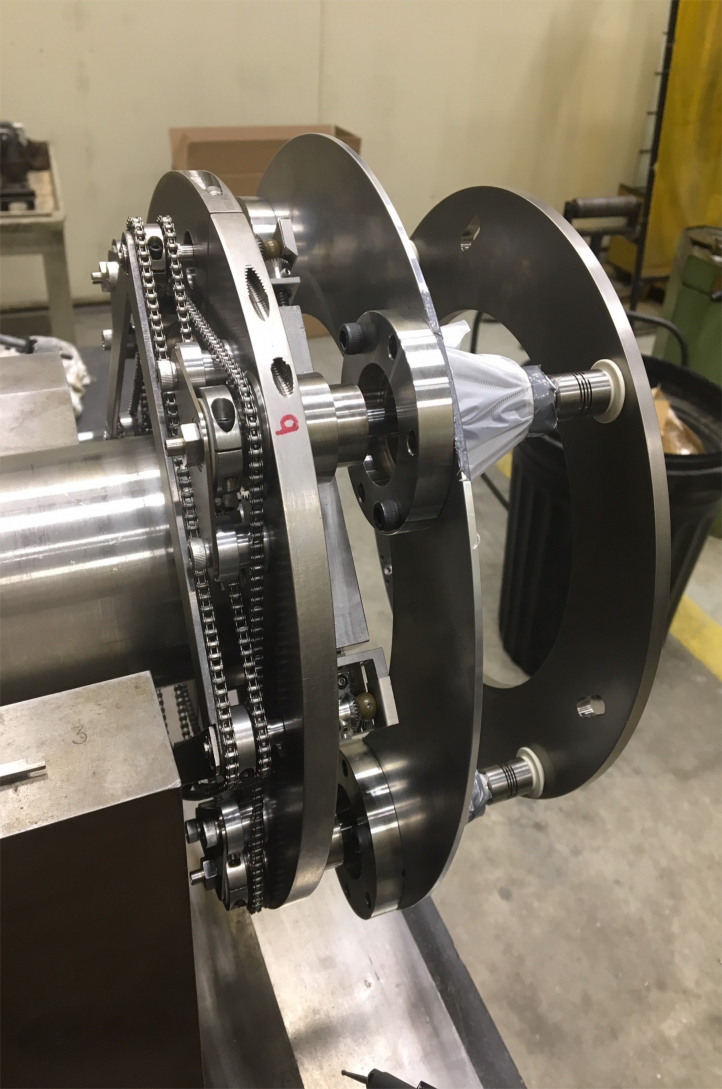}
\subcaption{view from downstream}\label{fig:1a}
\end{minipage}
\begin{minipage}[b]{.5\linewidth}
\centering
\includegraphics[width =0.75\textwidth]{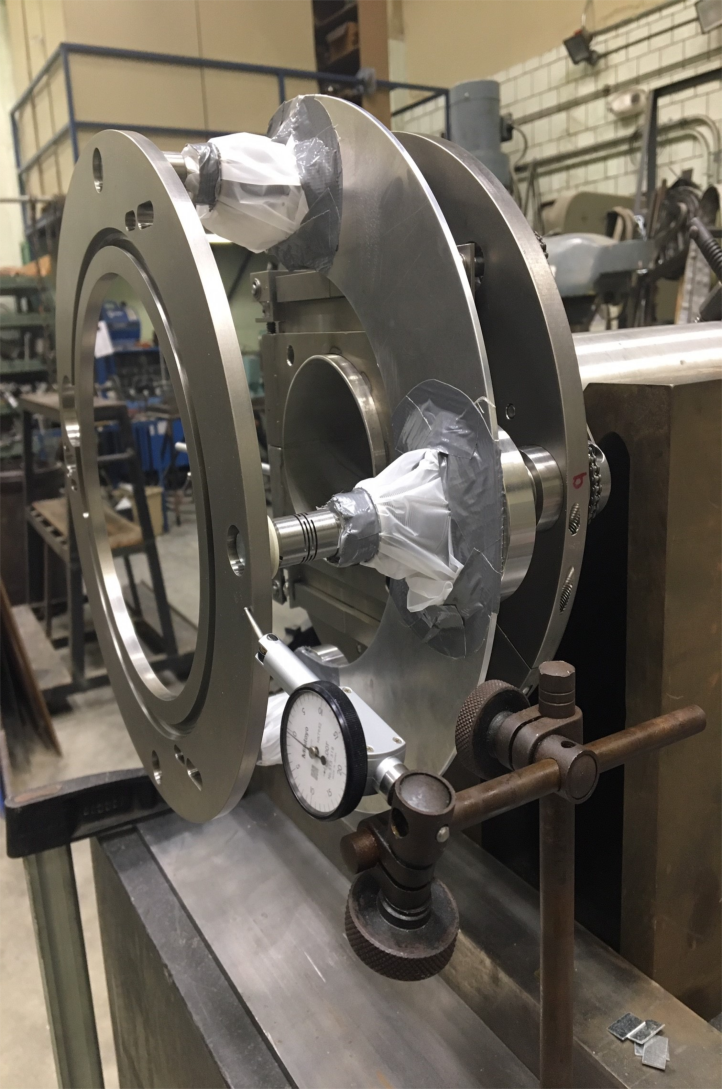}
\subcaption{view from upstream}\label{fig:1b}
\end{minipage}
\caption{View of the anode mounting assembly from different angles showing the mechanical arrangement for anode movement. The chain arrangement is responsible for the anode movement and is outside vacuum. The actual anode, to be installed on the leftmost circular plate on figure b, is not shown.}
\label{fig:anode}
\end{figure}

\begin{figure}[t!]
 \centering   
  \includegraphics[width=0.9\textwidth]{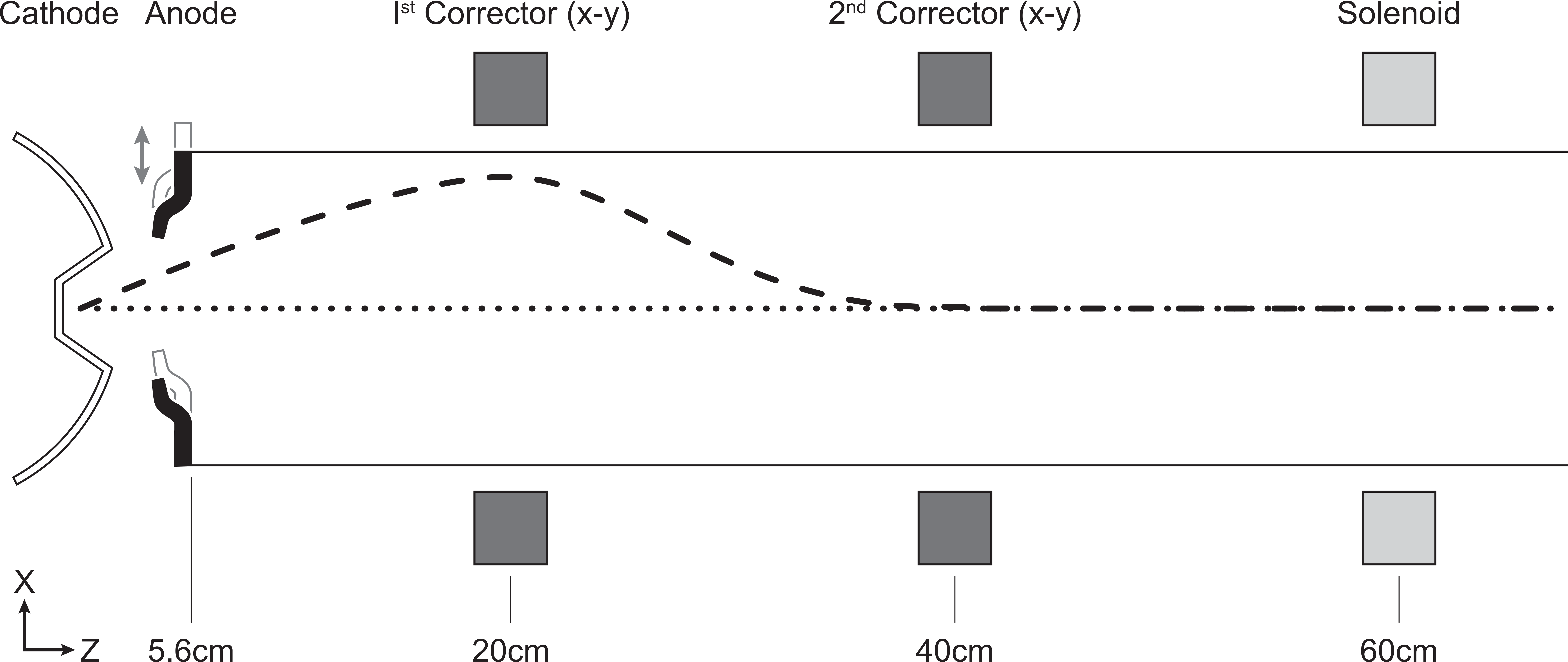}
    \caption{Schematic of the beam line components used for simulation, the Helmholtz coil in the DC gap is not shown. The dotted and dashed lines represents on axis and offset anode operation respectively. Components and distances are not to scale.}
    \label{fig:beamline}
\end{figure}

Figure \ref{fig:anode} shows different views of the anode mounting assembly during fabrication. The chain arrangement in figure \ref{fig:1a} contains actuators for accurate alignment between cathode and anode. These actuators also provide X-Y motion, pitch and yaw. The maximum displacement on each axis is 9mm in both directions. The chain assembly is outside vacuum whereas the leftmost circular plate (in figure \ref{fig:1b}) is inside vacuum where the anode gets installed. The middle circular plate in both figure \ref{fig:1a} and \ref{fig:1b} represents the point on the gun vessel where this assembly will be attached. The anode is electrically isolated and can be biased. The in-vacuum parts have been vacuum fired to achieve low outgassing rates.
 \par
Since the laser spot is at the center of the cathode, the beam will has to to be deflected towards the center of the anode as it is exiting the DC gap. A Helmholtz coil, located outside of the gun vacuum chamber, can be used to deflect the beam to make it pass through the center of the anode. After the beam has passed the DC gap, subsequent corrector coils and solenoids are used to steer the beam to on axis of the beam line and compensate for emittance growth. The recessed structure and the chain assembly restrict how close the first X-Y corrector can be installed. For this design, 20 cm downstream from the cathode is the closest distance for the first corrector magnet. The second set of correctors are at 40 cm and the solenoid is at 60 cm downstream. A schematic diagram of the beam line is shown in figure \ref{fig:beamline}. \par

\begin{figure}[h!]
\centering   
  \includegraphics[width=0.9\textwidth]{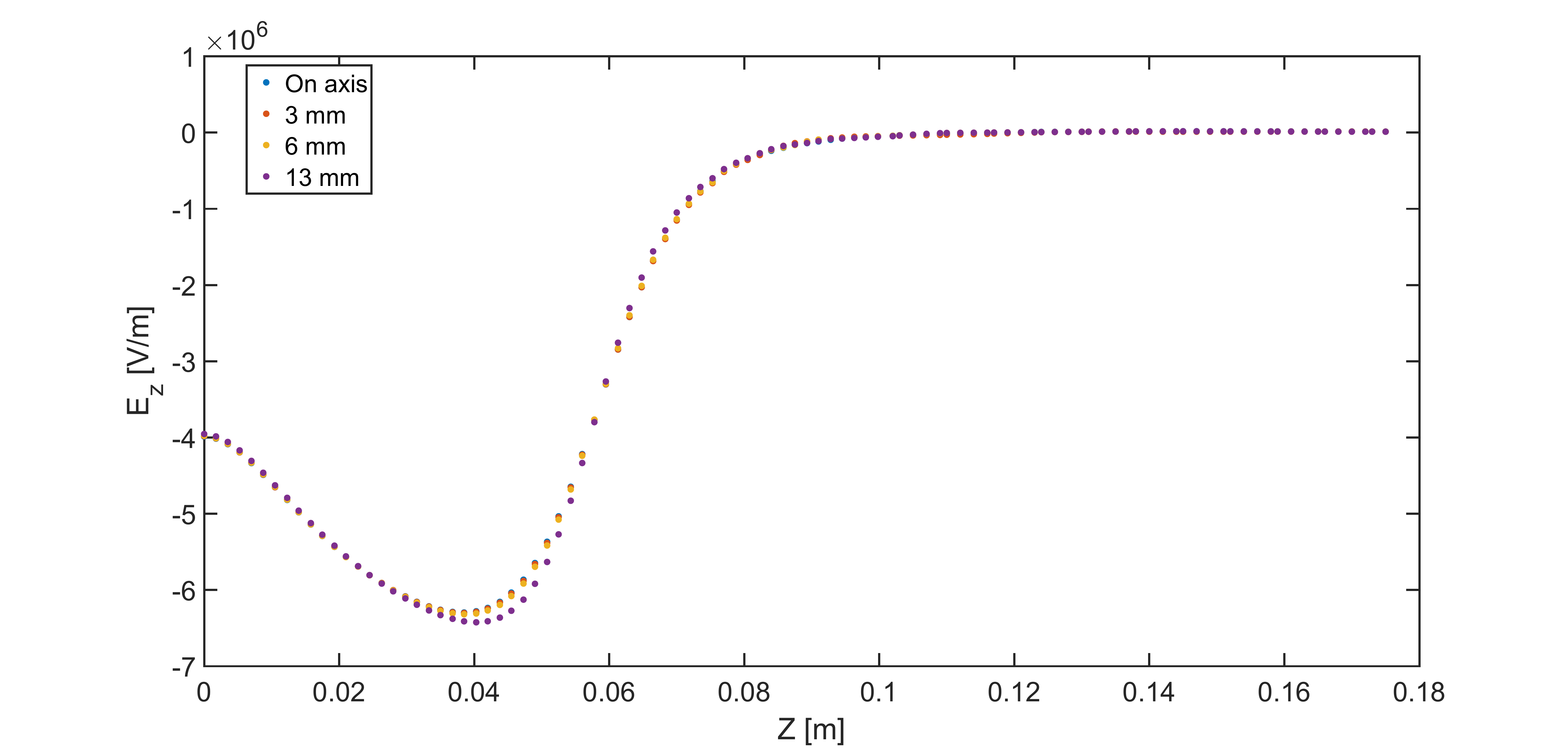}
    \caption{Electric field along Z for various anode offsets}
  \label{fig:Ez}
\end{figure}
We used Opera 3D to simulate the 3D electric field distribution in the DC gap for 4 different anode offset cases: on axis, 3 mm offset, 6 mm offset and 13 mm offset \cite{Opera}. All the displacements of the anode were in the positive X direction. The gradient on the center of the cathode is 4 MV/m and the E$_z$ profile remains unchanged for various anode offsets as shown in  figure \ref{fig:Ez}. The longitudinal field goes to zero at around 20 cm downstream from the cathode surface. The corrector magnets used in this simulation on the beam line are standard elements from GPT. The solenoid field used in the simulation was the measured profile from the fabricated solenoid.

\section{Ion back bombardment: Theory, simulation results and discussion}

\subsection{Theory of ion back bombardment}
Ion back bombardment rate, and consequent decay of cathode quality, is dependent on the vacuum parameters, current parameters, and voltage of the DC gun. At the beginning of the beam extraction, the entire laser spot area on the activated photocathode is capable of emitting electrons. As time progresses, the NEA layer deteriorates due to ion back bombardment and part of the laser illuminated area becomes incapable of electron emission. The decay rate of the emission area is directly proportional to the ion generation rate. If $A_e(0)$ is the emission area capable of emitting electrons at t = 0 and $A_e(t)$ is the emission area at time t, the differential equation describing the decay of the emission area can be written as,
\begin{equation}
    (\frac{A_e(t)}{A_e(0)}) (\frac{dN_i}{dt}) S dt = -dA_e(t)
    \label{eq:1}
\end{equation}
where $\frac{dN_i}{dt}$ is the ion generation rate due to electron beam passing the DC gap and $S$ is the damage coefficient of the generated ions to the surface. The severity of cathode damage is dependent on the ion's energy. The higher energy ions will penetrate into the crystal and will result in lattice distortion and back sputtering while low energy ions may sputter off the surface atoms of activation layers \cite{liu2016effects}. Experimental studies have been performed to understand the surface layer damage after beam delivery but there is no conclusive quantitative model to describe the damage rate as the function of ion energy \cite{shutthanandan2012surface}. Therefore for simplicity, we assume that the weight of ions energy damage effect ($S$) is one.\par
The ion generation rate is directly proportional to the electron beam current, the scattering cross section, residual gas pressure and the electric field gradient. It can be written as,
\begin{equation}
    \frac{dN_i}{dt} = I \frac{n_i}{e} \int_{0}^{d} \sigma(E_z) dz = I \times F
    \label{eq:4}
\end{equation}
where $I$ is the electron beam current, $\sigma$ is the ionization cross section of Hydrogen molecule, $E_z$ is the accelerating field, $n_i$ is density of residual gas molecule and $d$ is the DC gap distance.
When the emission current is constant, the solution to equation \ref{eq:1} is of exponential form,
\begin{equation}
    A_e(t) = A_e(0)e^{(\frac{-t}{\tau})}
    \label{eq:2}
\end{equation}
with the time constant, $\tau$, being related to available active area and incoming particle flux as,
\begin{equation}
    \tau = \frac{A_e(0)}{(\frac{dN_i}{dt})} = \frac{A_e(0)}{I \times F}
    \label{eq:3}
\end{equation}
Then equation \ref{eq:3} can be written as, 
\begin{equation}
    A_e(t) = A_e(0) e^{-\kappa Ft}
    \label{eq:5}
\end{equation}
With $\kappa$ relating the average current and initial emission capable area as,
\begin{equation}
    \kappa = \frac{I}{A_e(0)}
\end{equation}
\par
From equation \ref{eq:3}, it is evident that the time constant, hence the charge lifetime, will decrease if the average current is increased while keeping the initial emission area (equivalent of the laser spot size) fixed. In order to increase the charge lifetime for high average current, the laser spot size has to be increased accordingly. This is one of the major motivations behind the design of the BNL prototype gun, where the optimized laser spot size is determined to be 8.6 mm in diameter.\par 
The ionization cross section can be written as \cite{reiser2008},
 \begin{equation} \label{eq:6}
 \sigma [m^2] = \frac{1.301\times 10^{-24}}{\beta ^2}f(\beta)\Big[ln(1.177 \times 10^5\beta^2 \gamma^2) - \beta^2\Big]
 \end{equation}
 with,
 \begin{equation} \label{eq:7}
 f(\beta) = \frac{6.027\times 10^{-5}}{\beta^2}(1.659 \times 10^4\beta^2 -1)
 \end{equation}
\par
With the electric field distribution along the beam path and using equations \ref{eq:6} and \ref{eq:7}, the ion generation flux, $F$, can be calculated.
\begin{figure} [h!]
\centering
    \includegraphics[width =0.8\textwidth]{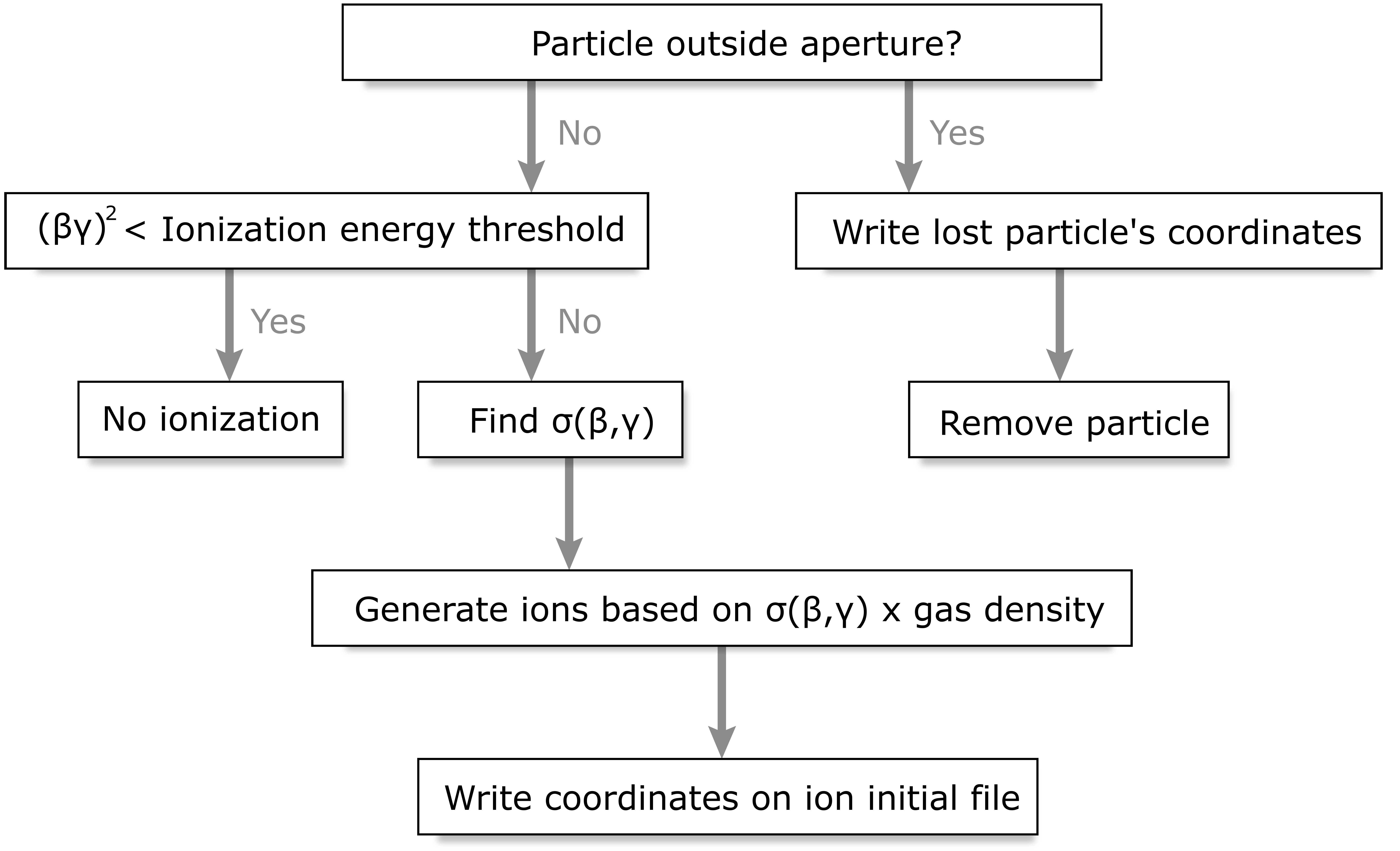}
    \caption{Flow chart describing the ion generation algorithm used in this study}
    \label{fig:ionflow}
\end{figure}
\par
We used "General Particle Tracer (GPT)" code to simulate both the electron beam and ion trajectories \cite{gpt05}. A custom GPT element was written to simulate the ion generation along the path of the beam. The source code for the element can be found at the URL in reference \cite{ibb}. Using equation \ref{eq:6} and \ref{eq:7}, the ionization cross section for each electron at every time step can be calculated as the electrons are passing the DC gap. The ion generation rate, which is linear with the product of cross section and gas density, can then be calculated using equation \ref{eq:4}. Figure \ref{fig:ionflow} shows the flow chart  for the ion generation algorithm. \par
The simulation was performed in two steps. First we simulated the electron beam, using the 3D electrostatic field profile obtained from Opera3D. The ions were also generated during this step, recording their initial positions as an initial ion distribution file. The initial hydrogen ion momentum distribution is a Maxwell distribution which is much slower than the kinetic energy from static electric field. Thus we assumed that the initial momentum of the generated ions to be zero. In the second step, the ions were accelerated using the initial ion distribution file in the same DC field. Since the trajectory of the electrons and ions are simulated separately, the effect of ion trapping in electron beam  downstream of the anode was not considered \cite{ion_trap}. In the DC gap, the trajectories of the electrons and ions for offset anode operations do not overlap entirely and therefore ion trapping in the DC gap can also be ignored. The leakage field through the anode aperture was considered and ions were generated up to approximately 20 cm downstream where the DC field goes to zero.

\begin{figure}
\begin{minipage}[b]{.5\linewidth}
\centering
\includegraphics[width =0.75\textwidth]{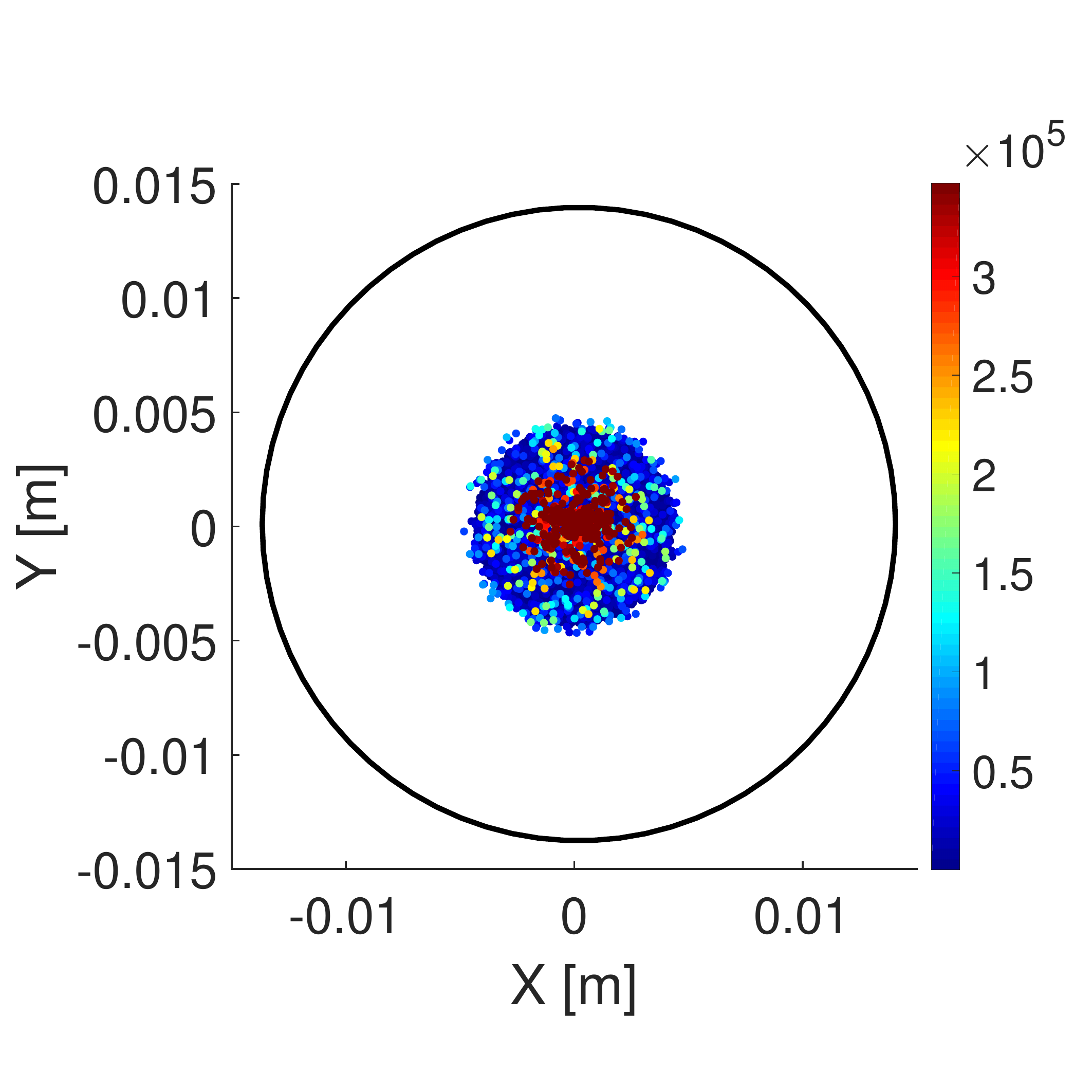}
\subcaption{On axis}\label{fig:5a}
\end{minipage}
\begin{minipage}[b]{.5\linewidth}
\centering
\includegraphics[width =0.75\textwidth]{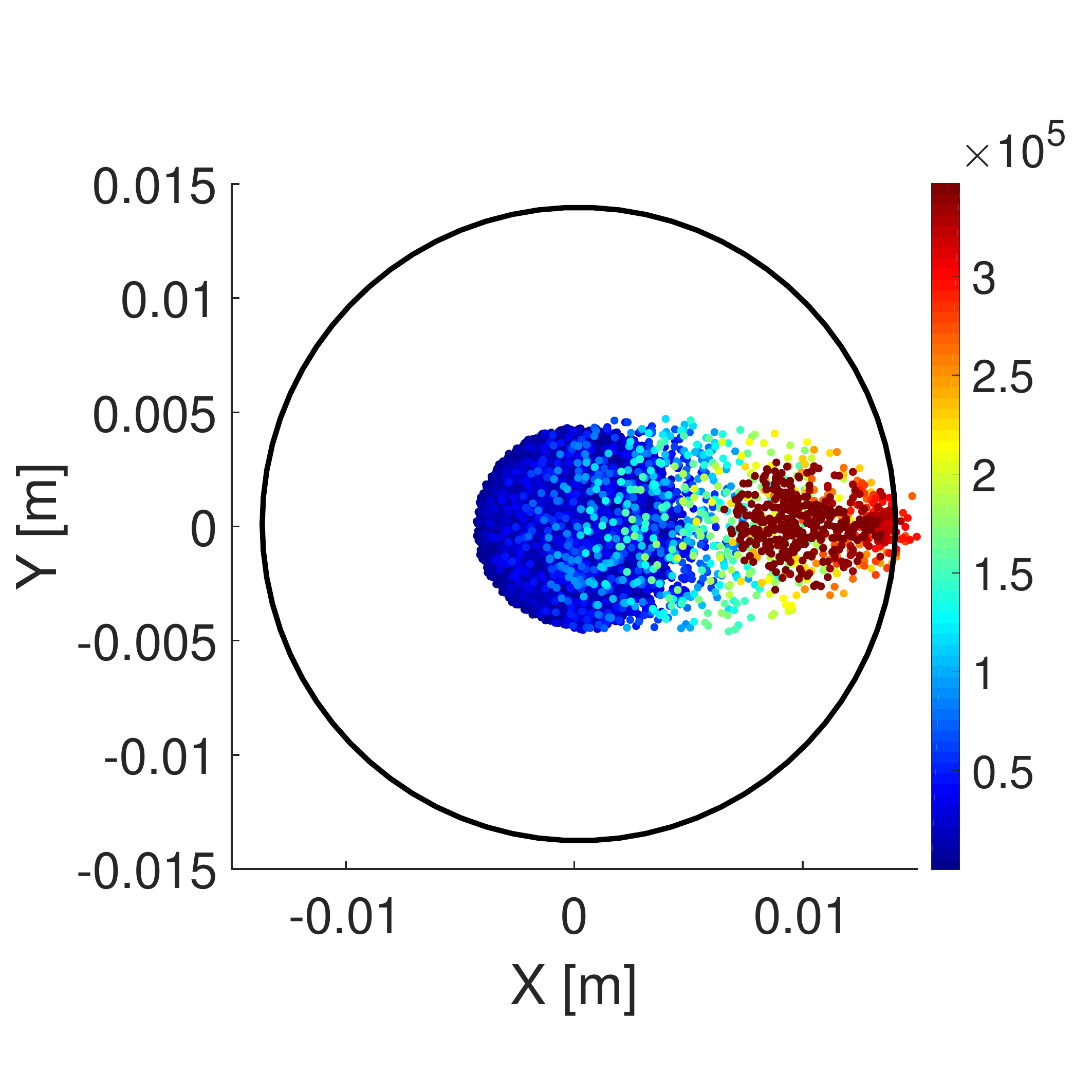}
\subcaption{Anode offset 13 mm}\label{fig:5b}
\end{minipage}
\begin{minipage}[b]{.5\linewidth}
\centering
\includegraphics[width =0.75\textwidth]{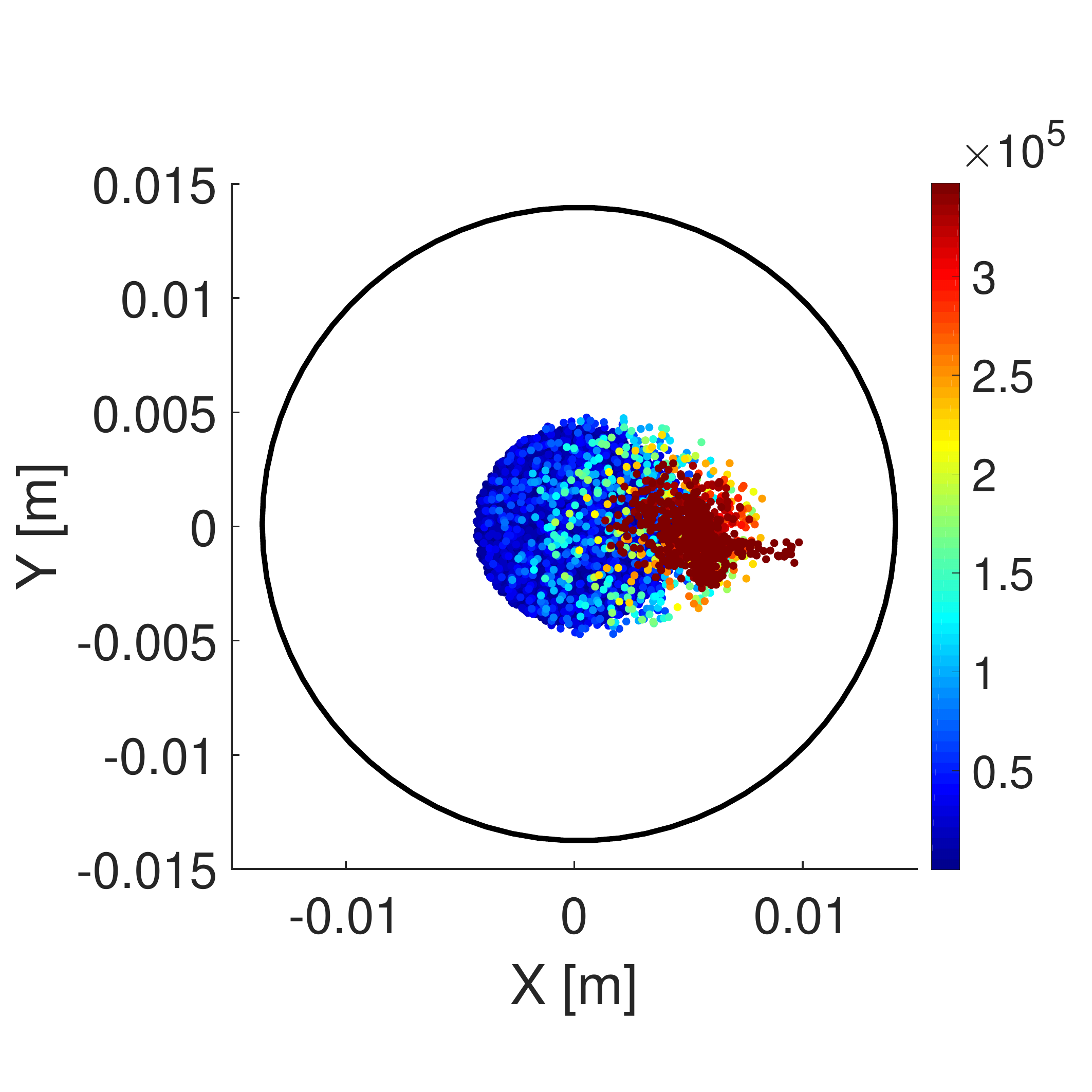}
\subcaption{Anode offset 6 mm}\label{fig:5c}
\end{minipage}
\begin{minipage}[b]{.5\linewidth}
\centering
\includegraphics[width =0.75\textwidth]{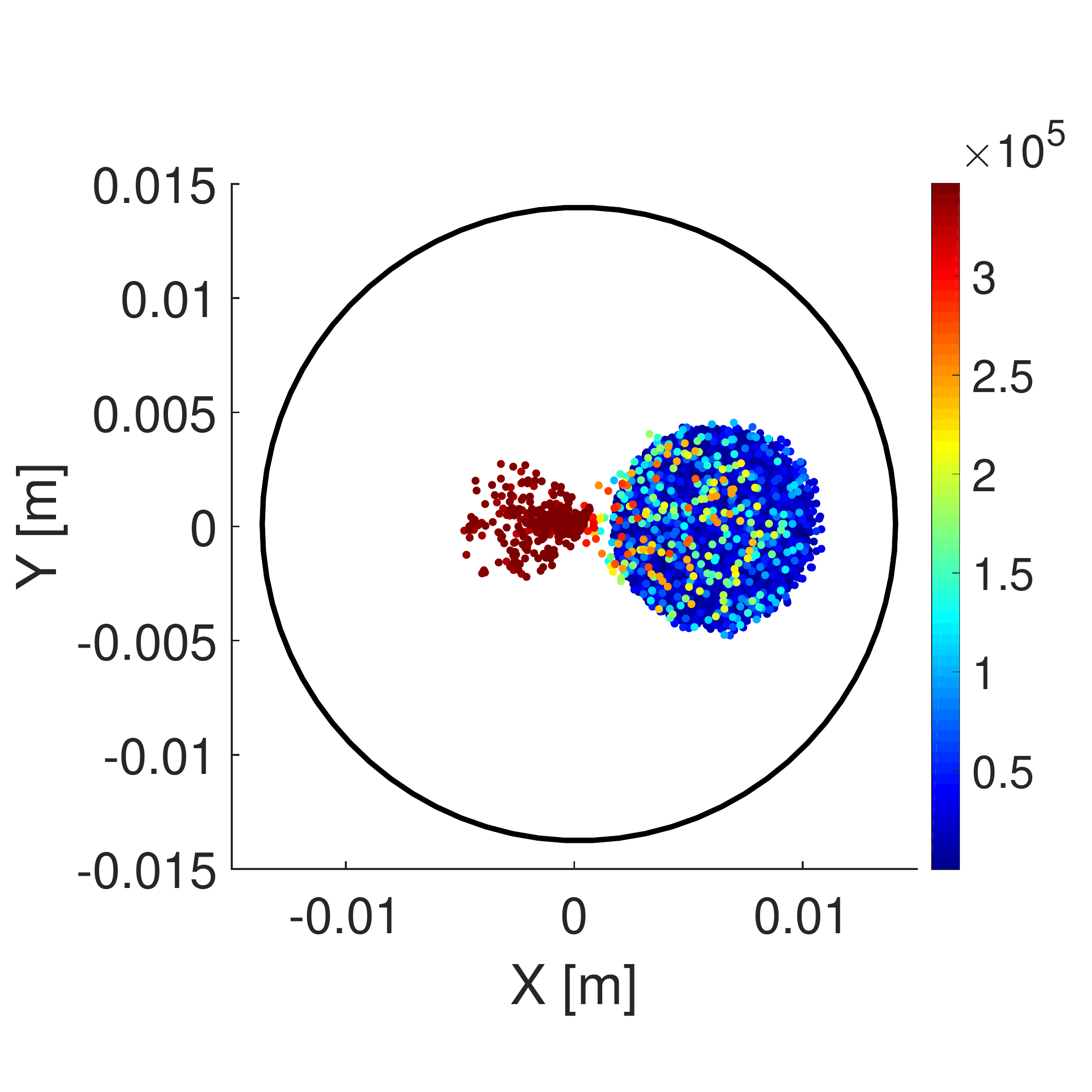}
\subcaption{laser offset 6 mm}\label{fig:5d}
\end{minipage}
\begin{minipage}[b]{.5\linewidth}
\centering
\includegraphics[width =0.75\textwidth]{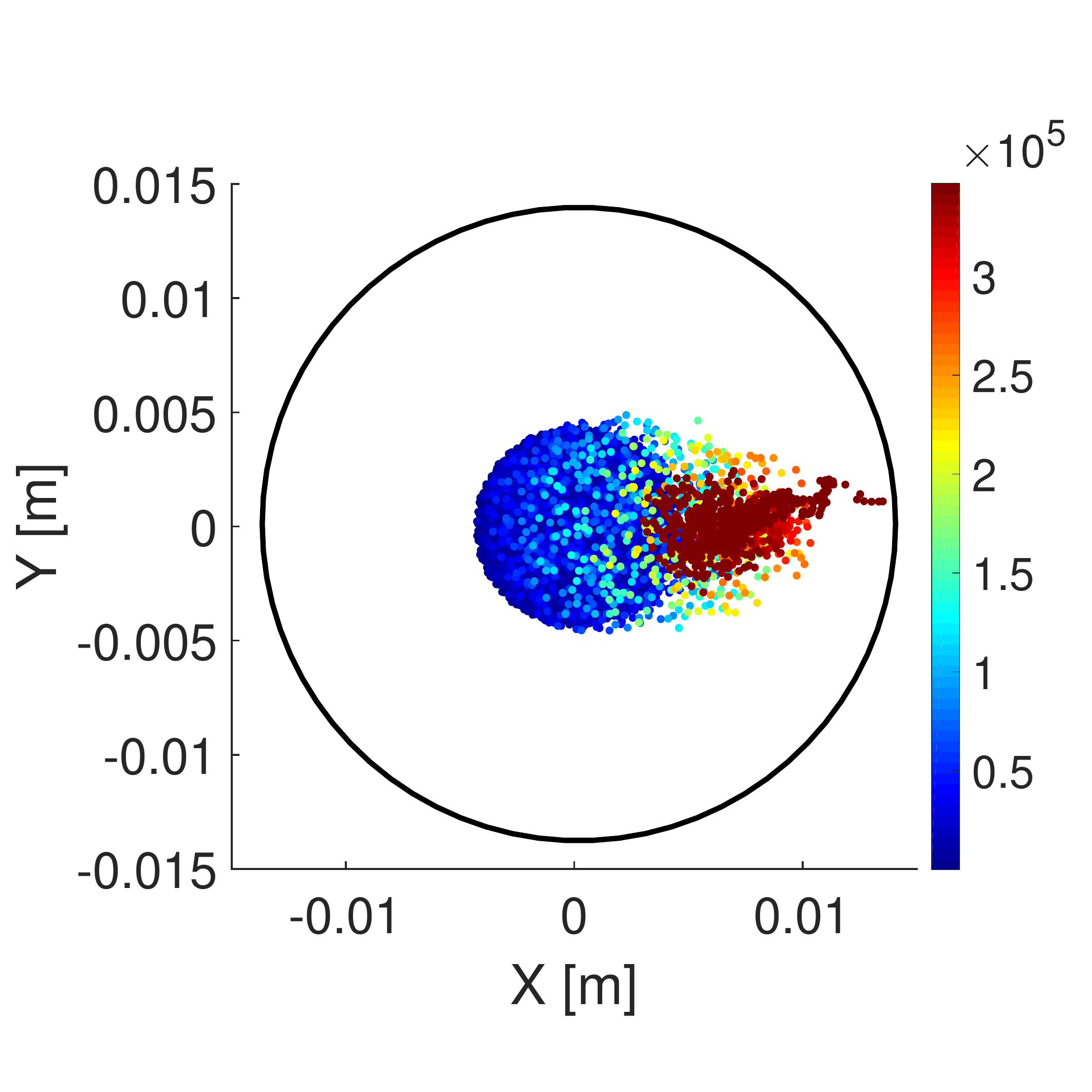}
\subcaption{Anode offset 8.5 mm}\label{fig:5e}
\end{minipage}
\begin{minipage}[b]{.5\linewidth}
\centering
\includegraphics[width =0.75\textwidth]{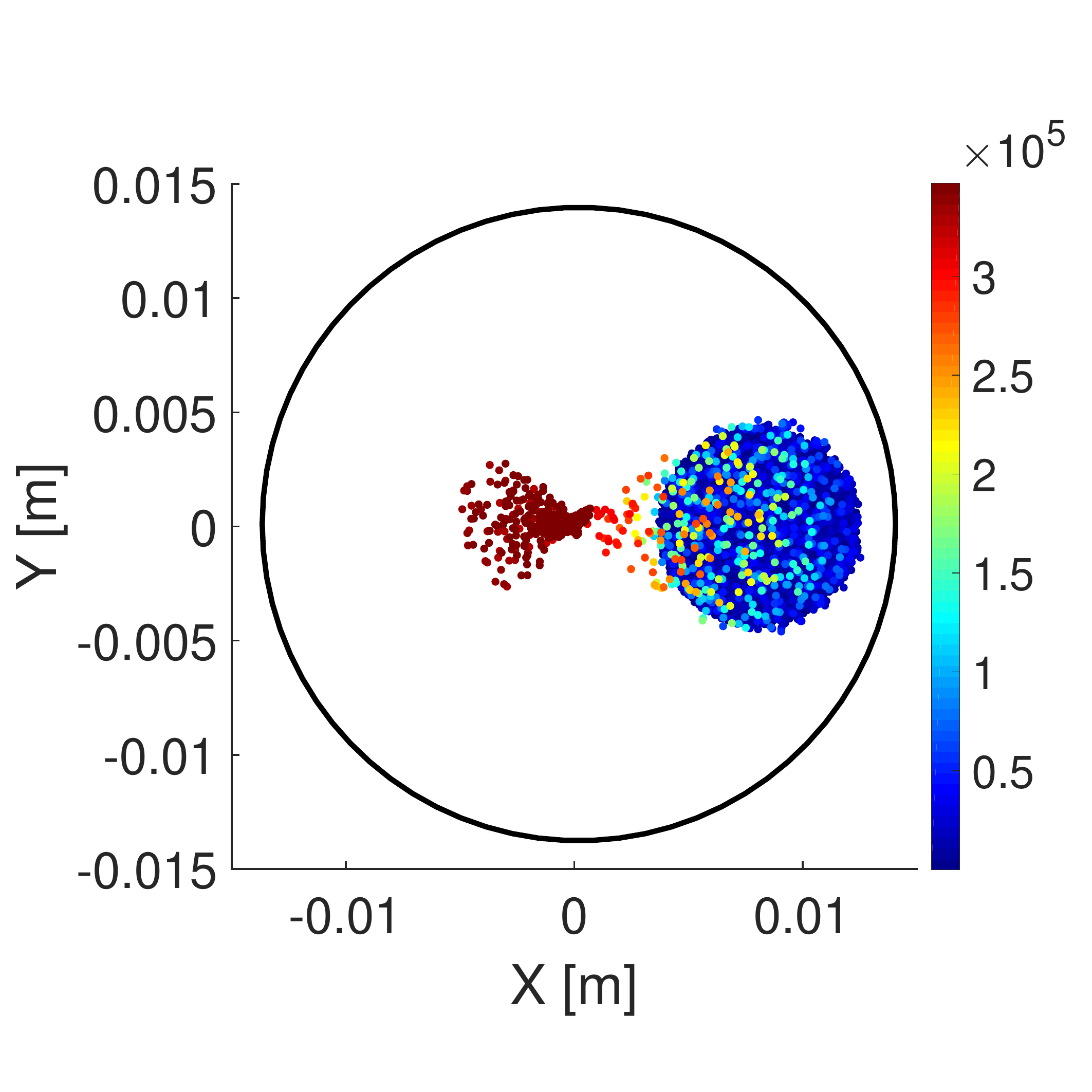}
\subcaption{laser offset 8.5 mm}\label{fig:5f}
\end{minipage}
\caption{Simulated ion distribution on the cathode. laser spot is 8.6mm in diameter. The colorbar represents the energy of the bombarding ions in eV.}
\label{fig:ion_map}
\end{figure}

\subsection{Ion back bombardment simulation results} \label{sec:5}
The ion maps on the cathode surface for various anode and laser offsets are shown in fig \ref{fig:ion_map}, with the colorbar representing the energy of the ions. The ionization cross section of Hydrogen due to electron-H$_2$ scattering is energy dependent and about 3 orders of magnitude higher for lower (less than 100 eV) energy compared to higher energy (more than 100 KeV) \cite{reiser2008}. Therefore most of the ions generated in the DC gap are very low energy ions from within the first few mm of the cathode surface. The ion maps from figure \ref{fig:ion_map} represents this with the high concentration of blue ions. For on axis operation, figure \ref{fig:5a}, high energy ions are focused on the center for on axis operation. This effect has been observed at Jlab \cite{jlab2007s}. Medium energy ions, represented by yellow and orange dots, are not as focused as the high energy ions and are seen to be present as far as the edge of the laser spot.  \par
\begin{figure}[tb!]
\begin{minipage}[b]{.5\linewidth}
\centering
\includegraphics[width =0.75\textwidth]{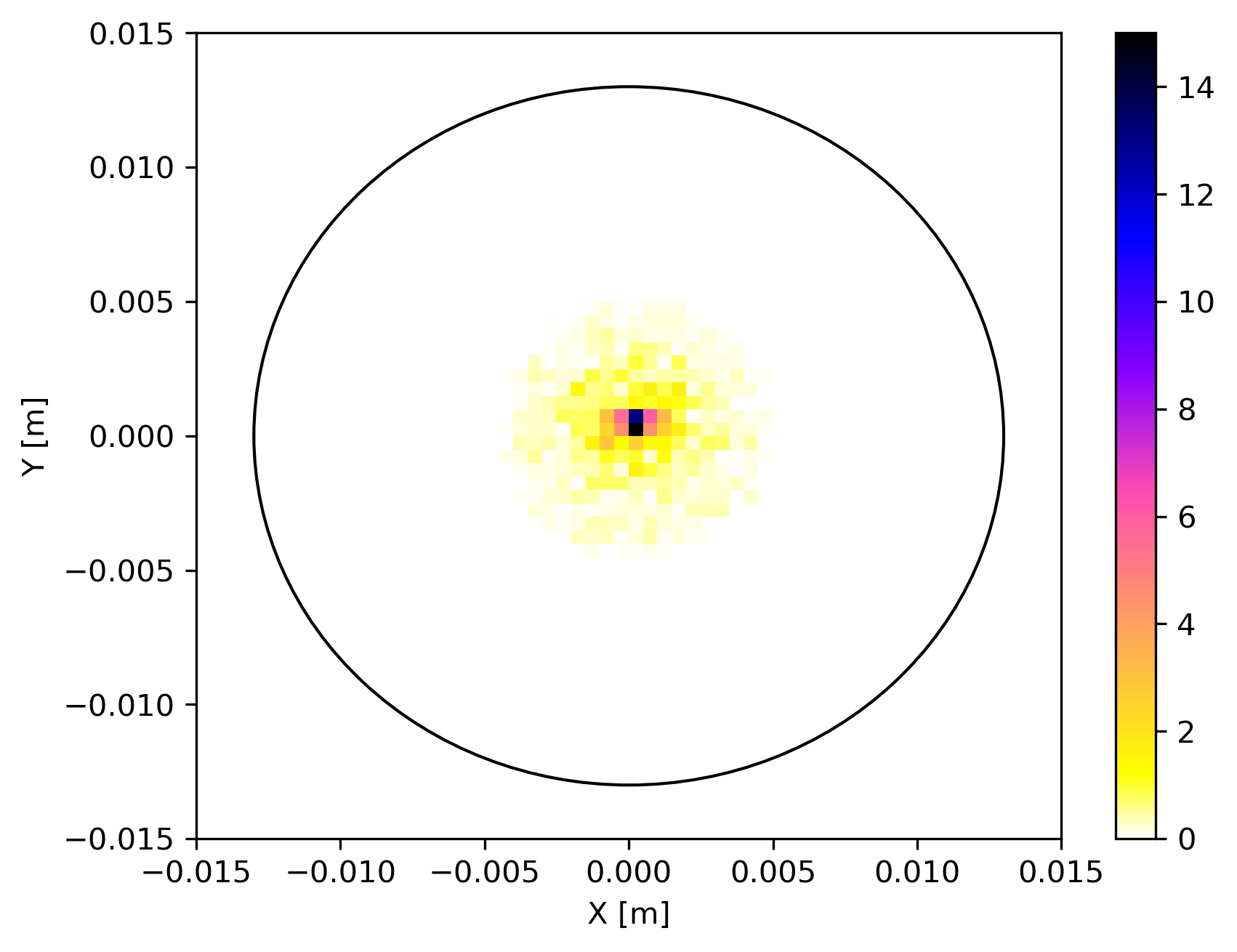}
\subcaption{On axis}\label{fig:6a}
\end{minipage}
\begin{minipage}[b]{.5\linewidth}
\centering
\includegraphics[width =0.75\textwidth]{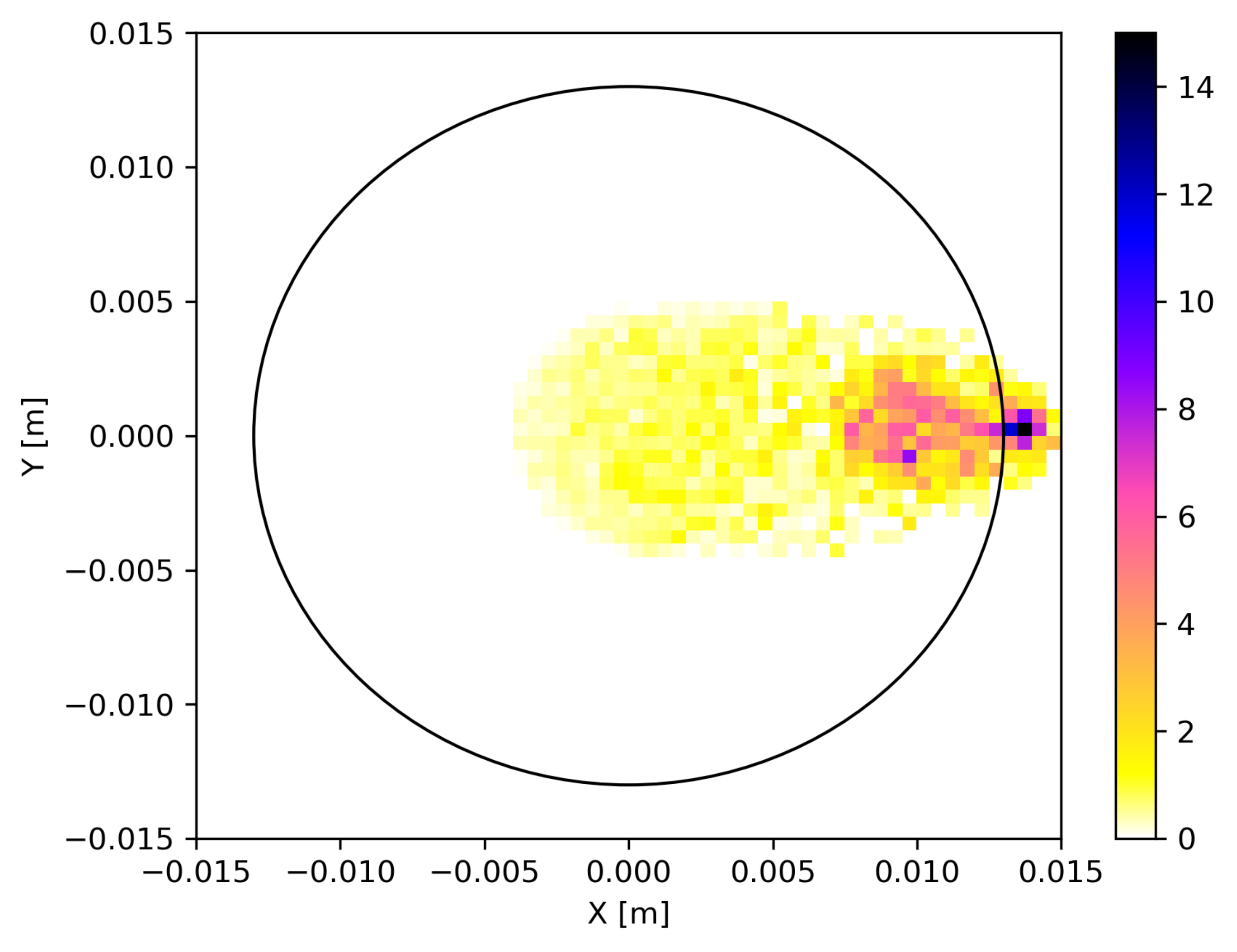}
\subcaption{Anode offset 13 mm}\label{fig:6b}
\end{minipage}
\begin{minipage}[b]{.5\linewidth}
\centering
\includegraphics[width =0.75\textwidth]{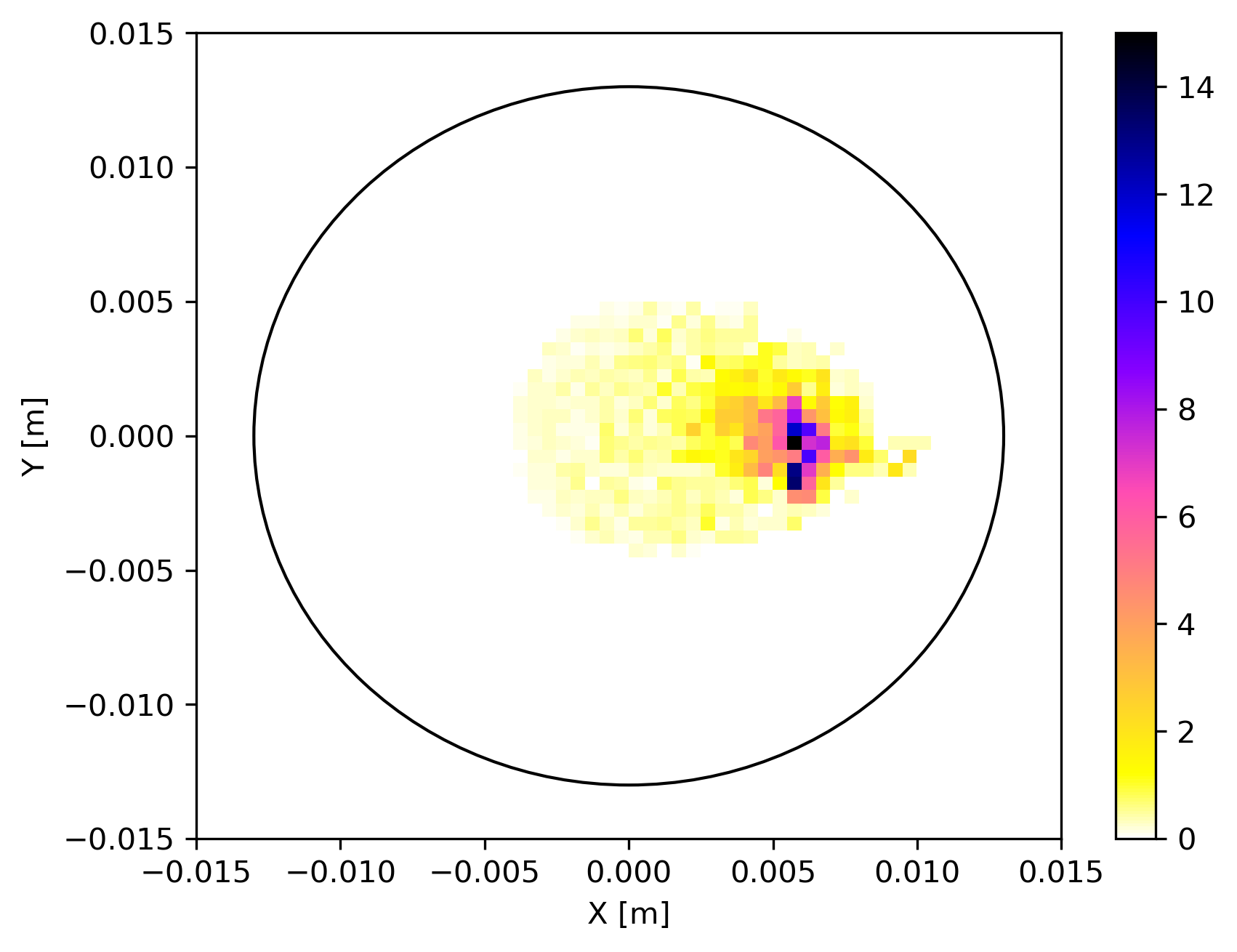}
\subcaption{Anode offset 6 mm}\label{fig:6a}
\end{minipage}
\begin{minipage}[b]{.5\linewidth}
\centering
\includegraphics[width =0.75\textwidth]{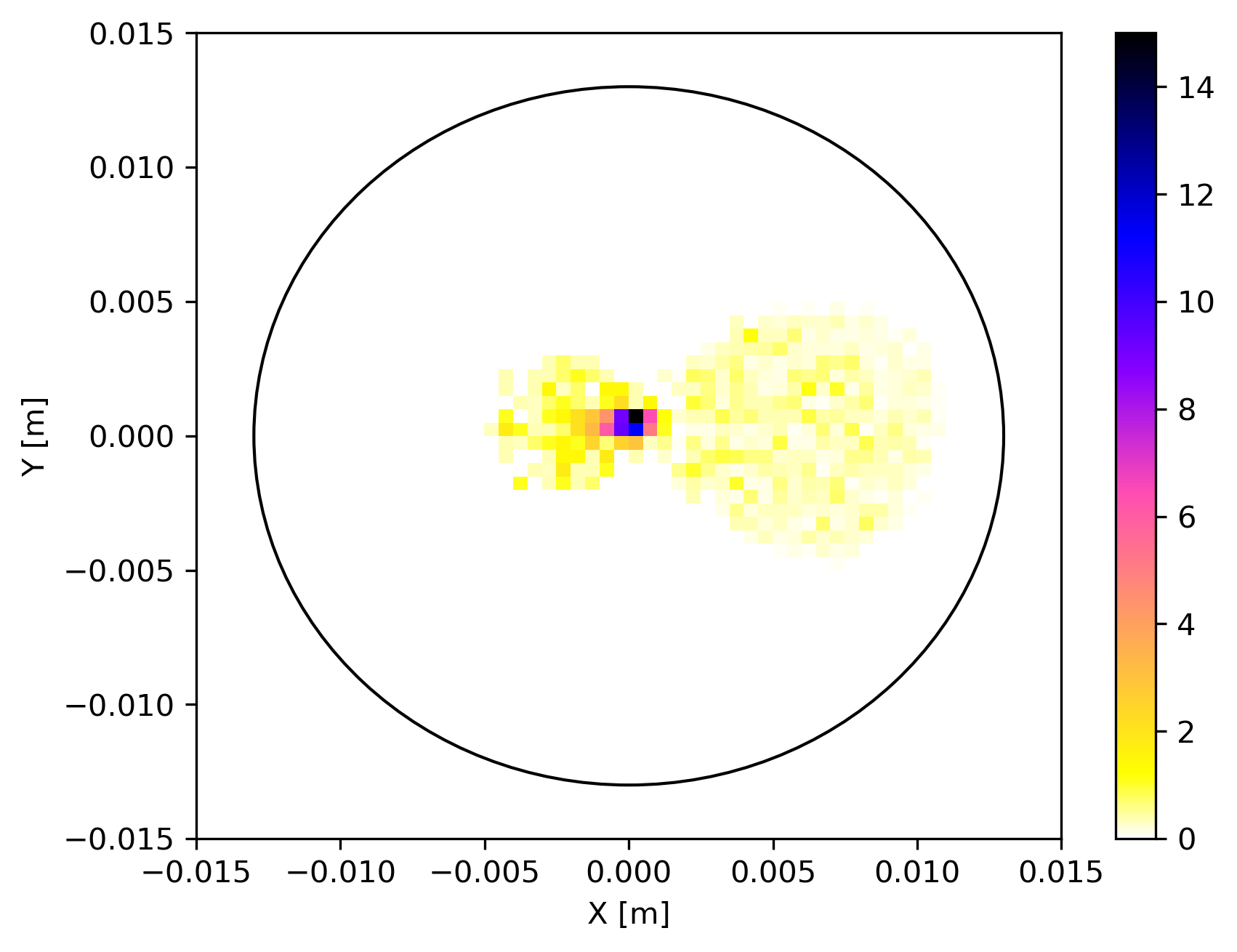}
\subcaption{Laser offset 6 mm}\label{fig:6b}
\end{minipage}
\begin{minipage}[b]{.5\linewidth}
\centering
\includegraphics[width =0.75\textwidth]{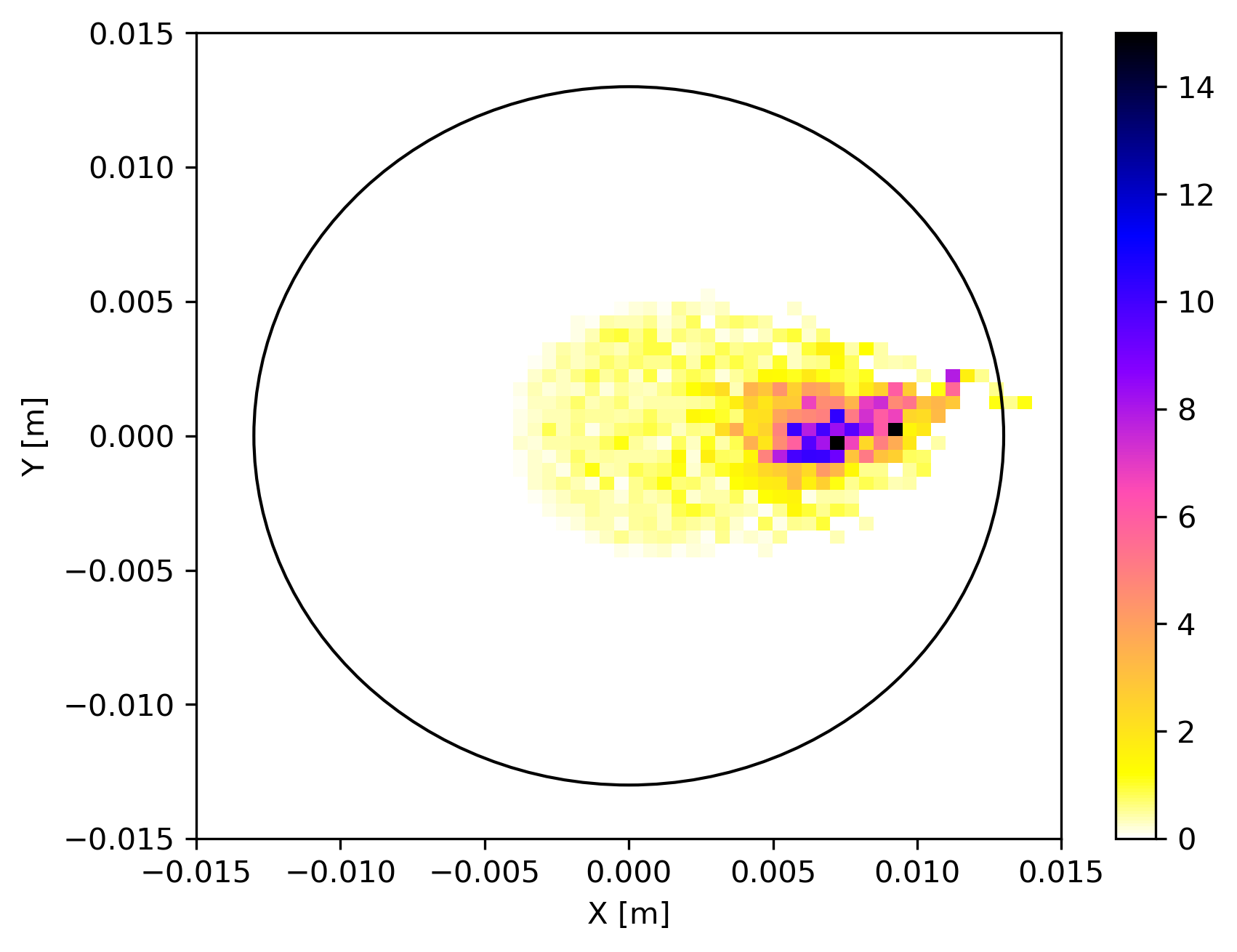}
\subcaption{Anode offset 8.5 mm}\label{fig:6c}
\end{minipage}
\begin{minipage}[b]{.5\linewidth}
\centering
\includegraphics[width =0.75\textwidth]{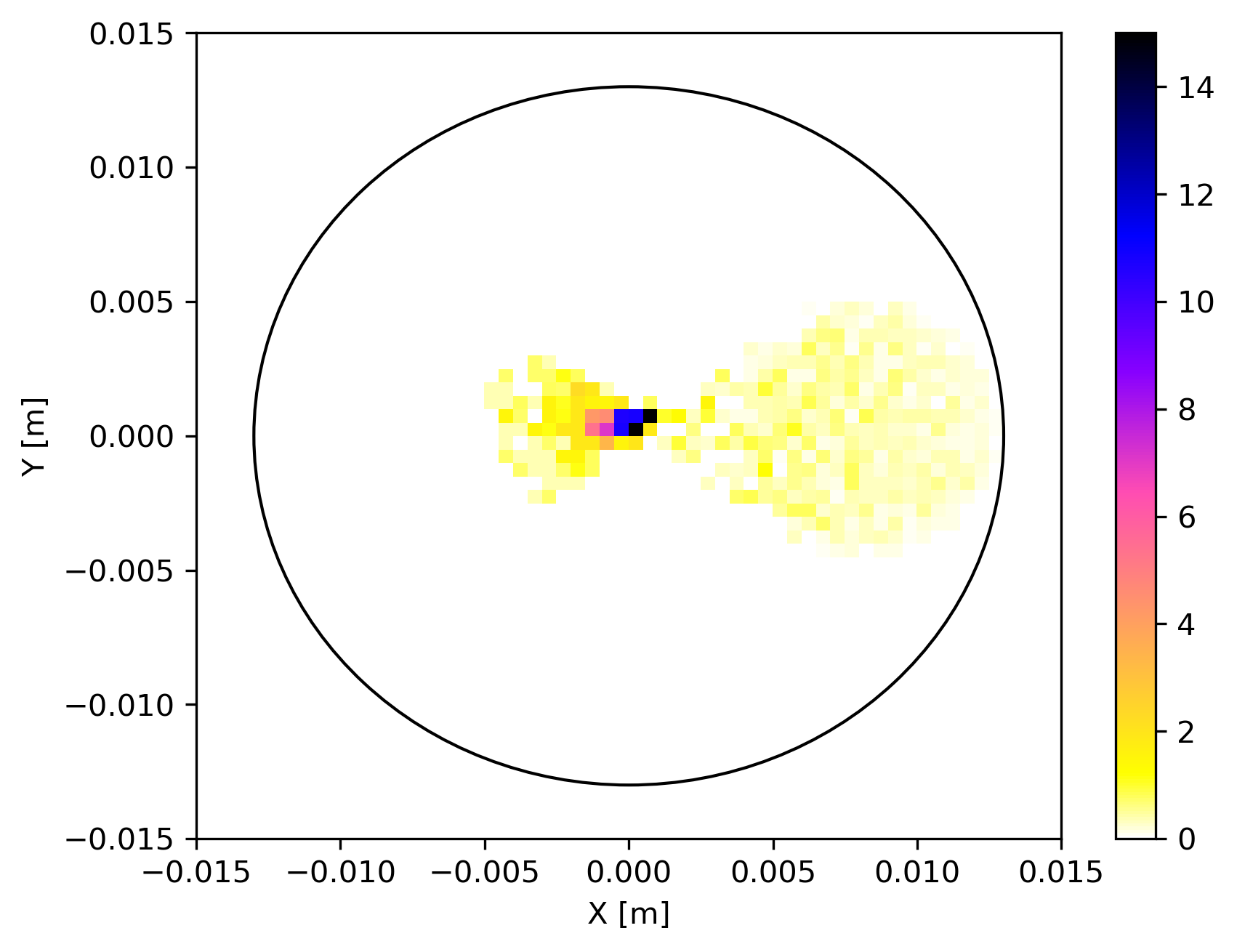}
\subcaption{Laser offset 8.5 mm}\label{fig:6d}
\end{minipage}
\caption{Power deposited on the cathode due to ion back bombardment for various anode and laser offsets. The power deposition is calculated on 0.5 mm X 0.5 mm grids and is normalized with maximum power deposited in one grid. The colorbar represents the normalized power deposited, white being no power and black being maximum power.}
\label{fig:pd}
\end{figure}
Figure \ref{fig:5d} and \ref{fig:5f} shows the ion map on the cathode for 6 mm and 8.5 mm laser offsets respectively. In both cases, the higher energy ions are focused at the electrostatic center. Some high energy ions are seen to be shifted in the opposite direction of the laser spot offset. These ions are generated when the beam is off axis on the DC gap - beam line and are then getting over focused in the DC gap due to the Pierce like focusing. The further the radial distance of the ion from the central axis, the more focusing it will experience in the DC gap. The fact that the electrostatic center is heavily bombarded by high energy ions for any amount of laser offset restricts the maximum size of the laser spot. One important feature from two offset laser ion maps is the area covered by the medium energy ions. In figure \ref{fig:5d}, the medium energy ions (yellow and orange dots) cover about 70\% of the laser spot, whereas in figure \ref{fig:5f} they cover about 50\% of the laser spot. This shows that with increasing laser offset, medium energy ion bombardment on the beam extraction spot will decrease. This should result in increasing charge lifetime with increasing laser offset for fixed laser spot. This effect was also experimentally observed at Jlab \cite{jlab2011}.
\par
For offset anode cases, the high energy ions are shifted towards the direction of the anode offset. With increasing anode offset, as seen in figures \ref{fig:5b}, \ref{fig:5c} and \ref{fig:5d}, the higher energy ions are seen to have moved outside the laser spot on the cathode. The shift distance for high energy ions is equivalent of the anode offset distance. We notice by comparing all the offset anode cases is that the higher energy ions get de-focused at the cathode with increasing anode offset. A closer inspection of the ion map for the 13 mm offset reveals that 350 KeV ions (dark red) are more de-focused than ions with energy between 300-350 KeV (light red). This is because these downstream ions are generated off center of the beam axis, on the positive quadrant of the XY plane. The E$_x$ field experienced by these ions, as they are traveling to the cathode, is negative. The further off axis the ion is, the higher the magnitude of the E$_x$ field, which results in the ions getting over-focused by the time they reach the cathode. These ions could be eliminated by simply biasing the anode. The ions that are generated close to the anode, but are in the DC gap, are not affected by this focusing as much. These ions follow a straighter trajectory and hit the cathode at a distance almost equal to anode offset. 
\par
The power deposited on the cathode was calculated on 0.5 mm by 0.5 mm grids using the energies of the bombarding ions for all anode and laser offset cases. The power per grid was normalized to the maximum power deposited on any grid and shown in figure \ref{fig:pd}. For on axis and laser offset operation, 90$\%$ of the total power is deposited within a 3-5 mm$^2$ area at the center of the cathode. For offset anode operation, the higher energy ions are more de-focused and shifted from the electrostatic center. For 8.5 mm and 6 mm anode offset, part of the laser spot is still affected by the higher energy ions. For 13 mm anode offset, the higher energy ions completely miss the laser spot at the center of the cathode.\par
It is noted that even for 13 mm, the medium-high energy ions do not completely miss the cathode area for this particular gun. In principle, we could offset the anode even further such that medium-high energy ions completely miss the cathode. However, this simulation study was performed keeping in mind the mechanical limitations of the BNL gun. We are limited in terms of the maximum anode offset because of the beam line aperture and large overall cathode size, as explained in the next section. Any offset larger than 13 mm for our case will require a 1st corrector magnet with cooling, which is not possible with the present design. If the cathode size is smaller than 26 mm, which is the case for most HV-DC polarized gun around the world, a much smaller offset would be able to shift the high energy ions completely. For example: if the total cathode diameter is 10 mm, a 7 mm anode offset would be adequate for shifting the high energy ions from the cathode surface.

\par
In reality, the vacuum in the DC gap is not uniform along the path of the beam and pressure profile could vary between the cathode surface to anode. To simplify the simulation, we only considered constant pressure profile in this simulation. However, the custom ionizer element can handle variable pressure profile as well.


\section{Beam dynamics simulation}\label{sec:4}

The beam parameters used for this study are the nominal parameters for the proposed BNL large cathode inverted gun. A 5.3 nC bunch with 1.2 ns bunch length and uniform density was simulated and propagated through the 2 m long beam line. The laser spot size is 8.6 mm in diameter, which is optimized for this particular gun design considering both cathode lifetime and beam quality. First we compare the beam quality for offset laser operation and offset anode operation. Then beam dynamics results for various offset anode operations are discussed. The beam dynamics study for offset anode operation focused on the trajectory of the beam (in particular the X trajectory), possible beam loss at various high risk spots on the beam line and emittance growth due to the transverse kick in the DC gap.
\par

\begin{figure} [h!]
\centering
    \includegraphics[width=0.8\textwidth]{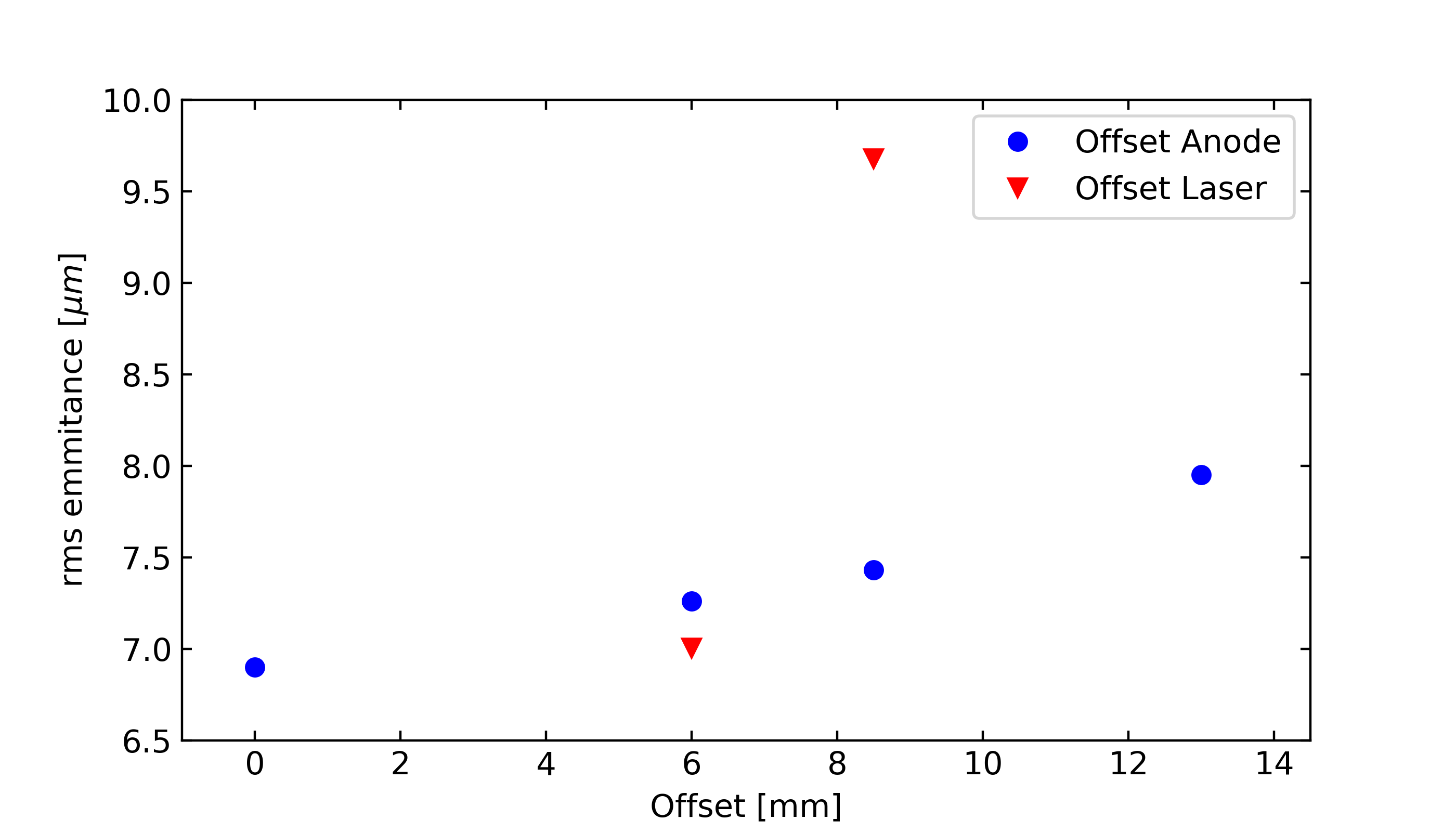}
    \caption{Normalized 4D rms emittances for all anode and laser offset cases after compensation by the solenoid.}
    \label{fig:exy}
\end{figure}
For the given laser spot size and overall cathode size in this gun, the maximum laser offset could be 8.5 mm. Figure \ref{fig:exy} compares the normalized 4D rms emittances for all laser and anode offset cases after compensation by the solenoid. For the 8.5 mm laser offset case, the emittance is found to be 9.68 $\mu$m, compared to 7.95 $\mu$m for 13 mm offset anode case. With increasing anode offset, the emittance seem to increase marginally which is expected. The 6 mm laser offset case emittance is almost equal to on axis operation case since this gun was specifically designed to tolerate that much laser offset. For 8.5 mm laser offset, the beam is almost approaching the edge of the cathode. This introduces non-linearities due to external field which dominated the emittance growth in this case. This was verified by comparing the emittance growth with and without space charge for various cases. This emittance growth due to non-linear external field could not be compensated using a solenoid. \par
Emittance growth due to geometric error or DC field aberration in the DC gun, for a small 0.35 mm laser spot, has been explored in reference \cite{bazarov2011comparison}. In the presence of linear focusing and space charge, the transverse momentum P$_x$ can be fitted at the exit of the gun as,
\begin{equation}
    P_x = Cx + \alpha_1 x^3
    \label{eq:px}
\end{equation}
Where C is the focusing constant and $\alpha_1$ is the aberration constant. The normalized transverse emittance is defined as
\begin{equation}
    \epsilon_{nx} =\frac{1}{mc}\sqrt{<x^2><P_x ^2> - <xP_x>^2} 
\label{eq:ex}
\end{equation}
Using equations \ref{eq:px} and \ref{eq:ex}, the normalized emittance at the anode is calculated to be,
\begin{equation}
    \epsilon_{nx} = \kappa \sigma_x ^4 \frac{\alpha_1}{mc}
    \label{eq:ex1}
\end{equation}
where $\kappa$ is a constant that depends on the particle distribution function. For our particular case, equations \ref{eq:ex} and \ref{eq:ex1} need to be revisited since the our laser spot size is an order of magnitude larger. The goal of the BNL gun is to have a longer charge lifetime, for which larger laser spot size is necessary. If the laser spot is large enough, for example 8.6 mm in our case compared to 0.35 mm in reference \cite{bazarov2011comparison}, operating the laser close to the edge of the cathode will introduce higher order non-linearity from the transverse field. Therefore the calculation leading to equation \ref{eq:ex1} has to be expanded to include higher order non-linearity coming from the transverse field.\par
The transverse non-linearity in the electric field will introduce a higher order term in P$_x$ such as,
\begin{equation}
    P_x = Cx + \alpha_1 x^3 + \alpha_2 x^n
\end{equation}
where n is an integer such that n$>$3 and $\alpha_2$ is the coefficient of the external field non-linear aberration. P$_x$ could be written in terms of a multiple higher order terms as well. For simplicity, we considered the most dominant non-linear term for this calculation.
Using the same method of calculation as in \cite{bazarov2011comparison}, we obtain the following formulas for the square of the normalized emittance,
\begin{equation}
\begin{aligned}
    \epsilon_{nx} ^2 = \frac{1}{mc} \left(\kappa_1 \alpha_1 ^2 \sigma_x ^8 + \kappa_2 \alpha_2 ^2 \sigma_x ^{2(n+1)} \right)  &    \hspace{2cm} \text{for n = even}\\
    \end{aligned}
    \label{eq:evenex}
\end{equation}
\begin{equation}
\begin{aligned}
    \epsilon_{nx} ^2 = \frac{1}{mc} \left(\kappa_1 \alpha_1 ^2 \sigma_x ^8 + \kappa_2 \alpha_2 ^2 \sigma_x ^{2(n+1)} + 2\alpha_1\alpha_2\sigma_x ^{(n+5)}\right)  &    \hspace{1cm} \text{for n = odd}\\
    \end{aligned}
    \label{eq:oddex}
\end{equation}
From the above equations it is clear that if the higher order non-linear term is dominant, i.e. $\frac{\alpha_1}{\alpha_2}<<1$, the emittance growth will be dominated by the non-linear term. If the non-linearity leads to severe distortion of  transverse phase space, it might not be feasible to fit P$_x$ as a function of x. 
In the case of 8.5 mm laser offset in our simulation, fitting x in terms of P$_x$, we get,
\begin{equation}
    x = CP_x + \alpha_1 P_x ^3 + \alpha_2 P_x ^4
\end{equation} where the ratio of $\alpha_1 / \alpha_2$ is of the order 1e-5. Therefore the x$^4$ term will dominate the emittance growth compared to the space charge related growth for 8.5 mm laser offset, which was seen from our simulation results. For the offset anode, the non-linearity introduced at the exit of anode is minimal. 

\begin{figure} [h!]
\centering
    \includegraphics[width=0.9\textwidth]{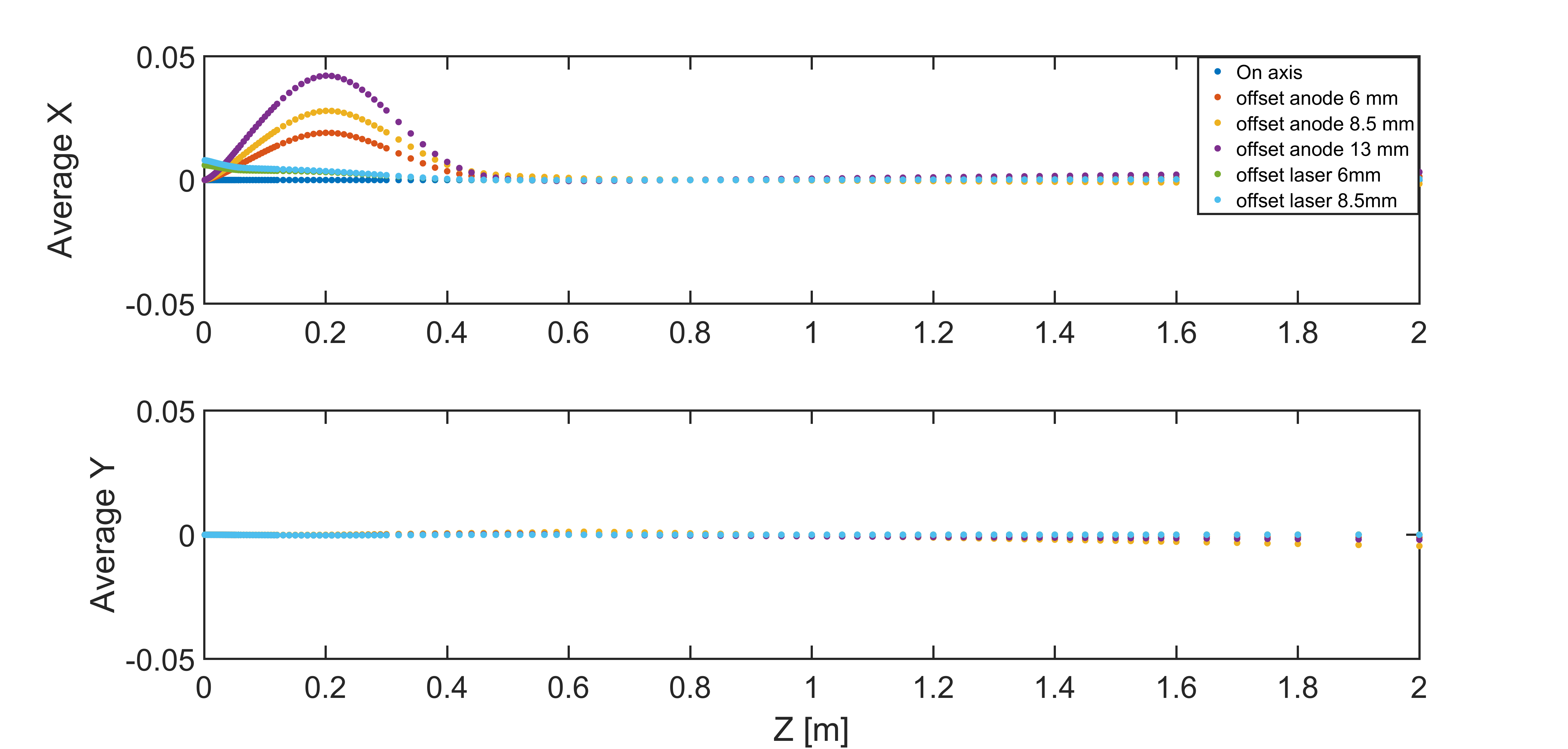}
    \caption{Trajectory of the bunch center in the X and Y direction along the beam line}
    \label{fig:avgx}
\end{figure}
\par
Figure \ref{fig:avgx} shows the X and Y trajectory of the center of the beam along the beam line for various anode offsets. Each point on the graph represents a 'screen' at that point on the beam line. The maximum deviation on the X axis is at the center of the 1st corrector magnet, which is expected. The potential of beam loss is maximum at the exit of anode (at 5.6 cm downstream), 20 cm downstream which is the center of the first X corrector and the entrance of the solenoid at 60 cm downstream.  Beam diameters in the X and Y direction at these to points are listed in table \ref{table:beamdia}. In all cases of anode offset, the transverse kick in the DC gap was set such that the beam center passes through the center of the anode, albeit at an angle. The position of the 1st X corrector magnet was fixed due to mechanical constraint and 20 cm is the closest possible location for this magnet. Comparing the beam diameters at the exit of anode for all anode offsets, it is clear that there should not be any beam loss at the anode since the anode aperture is 3.6 cm in diameter. The beam size in all cases are more than 2 times smaller than the anode aperture. At 20 cm downstream, beam has enough clearance considering the 10 cm diameter beam pipe for all the cases except 13 mm deviation. For this particular case, beam is within 10 mm of the beam pipe wall and could potentially cause major beam loss. This restriction limited the maximum anode offset to 13 mm for this particular gun setup. \par

\begin{table}[h]
\centering
\begin{tabular}{||c | c | c | c ||} 
 \hline
 Anode offset & Diameter(X/Y) at 5.6 cm & Diameter(X/Y) at 20 cm  & Diameter(X/Y) at 60 cm \\ [0.5ex]
 \hline
 mm & mm & mm & mm \\
 \hline\hline
 0 & 7.8/7.8 & 14/14 & 36/36 \\ 
 \hline
 3 & 7.7/7.8 & 14/14 & 36/36 \\
 \hline
 6 & 7.2/7.8 & 14/14 & 33/34\\
 \hline
 13 & 7/7.8 & 12/14 & 27/30 \\ [1ex] 
 \hline\hline
\end{tabular}
\caption{Beam size at the exit of anode (5.6 cm), at the center of 1st X corrector magnet (20 cm) and  at the entrance of the solenoid (60 cm) downstream for various offset anode operations. }
\label{table:beamdia}
\end{table}

From table \ref{table:beamdia}, it is also evident that with increasing anode offset, the beam will experience a focusing in the X direction as it passes through the DC gap due to the high transverse deflecting field. The mismatch in the sizes is partially recovered after the first solenoid. A quadrupole downstream could be used to match the beam sizes. The X-Y emittance mismatch could be compensated by the proper positioning of a solenoid.\par
\section{Conclusion}
We proposed and showed simulation results for an offset anode scheme for DC polarized electron guns that can substantially increase the charge lifetime from SL GaAs photocathodes. In this scheme, the effect of the high energy ion back bombardment is negated by essentially shifting these ions away from the cathode surface. This will ensure higher fluence lifetime and larger laser spot sizes compared to best alternative methods. The increment in charge lifetime can be a factor of ten or more depending on the overall size of the cathode.  The simulation results prove the feasibility of the scheme.
Beam dynamics results showed that a 5.2 nC bunch could be transported out of the gun with 20\% increase in normalized rms emittance for 13 mm anode offset compared on axis operation, which is acceptable. We have designed and fabricated necessary components, including an anode capable of in-vacuum movement, to perform charge lifetime measurements with different anode offsets.

\section*{Acknowledgements}
The authors are grateful to Dr. Ferdinand Willeke, Dr. Jorg Kewisch from BNL and Dr. Bruce Dunham from SLAC for useful discussions. The work was supported by U.S. Department of Energy under contract no DE-AC02-98CH10886.
\bibliography{IBB_draft_1.bbl}

\begin{thebibliography}{10}
\expandafter\ifx\csname url\endcsname\relax
  \def\url#1{\texttt{#1}}\fi
\expandafter\ifx\csname urlprefix\endcsname\relax\def\urlprefix{URL }\fi
\expandafter\ifx\csname href\endcsname\relax
  \def\href#1#2{#2} \def\path#1{#1}\fi

\bibitem{alley1995stanford}
R.~Alley, M.~Zolotorev, L.~Klaisner, C.~Garden, G.~Mulhollan, J.~Clendenin,
  H.~Tang, K.~Witte, E.~Hoyt, D.~Schultz, et~al., The stanford linear
  accelerator polarized electron source, Tech. rep., SCAN-9506065 (1995).

\bibitem{cates1989bates}
G.~Cates, V.~Hughes, R.~Michaels, H.~Schaefer, T.~Gay, M.~Lubell, R.~Wilson,
  G.~Dodson, K.~Dow, S.~Kowalski, et~al., The bates polarized electron source,
  Nuclear Instruments and Methods in Physics Research Section A: Accelerators,
  Spectrometers, Detectors and Associated Equipment 278~(2) (1989) 293--317.

\bibitem{hartmann1990source}
W.~Hartmann, D.~Conrath, W.~Gasteyer, H.~Gessinger, W.~Heil, H.~Kessler,
  L.~Koch, E.~Reichert, H.~Andresen, T.~Kettner, et~al., A source of polarized
  electrons based on photoemission of gaasp., Nuclear Instruments \& Methods in
  Physics Research Section A-Accelerators Spectrometers Detectors and
  Associated Equipment 286~(1-2) (1990) 1--8.

\bibitem{jlab2007s}
C.~K. Sinclair, P.~A. Adderley, B.~M. Dunham, J.~C. Hansknecht, P.~Hartmann,
  M.~Poelker, J.~S. Price, P.~M. Rutt, W.~J. Schneider, M.~Steigerwald,
  Development of a high average current polarized electron source with long
  cathode operational lifetime, Phys. Rev. ST Accel. Beams 10 (2007) 023501.

\bibitem{nishitani2005highly}
T.~Nishitani, T.~Nakanishi, M.~Yamamoto, S.~Okumi, F.~Furuta, M.~Miyamoto,
  M.~Kuwahara, N.~Yamamoto, K.~Naniwa, O.~Watanabe, et~al., Highly polarized
  electrons from gaas--gaasp and ingaas--algaas strained-layer superlattice
  photocathodes, Journal of applied physics 97~(9) (2005) 094907.

\bibitem{sinclair1999recent}
C.~Sinclair, Recent advances in polarized electron sources, in: Particle
  Accelerator Conference, 1999. Proceedings of the 1999, Vol.~1, IEEE, 1999,
  pp. 65--69.

\bibitem{sinclair2001}
C.~Sinclair, in: Snowmass 2001. Report No. SLAC-R-599, 2001.
\newblock \href{http://
  www.slac.stanford.edu/econf/C010630/proceedings.shtm}{[link]}.
\newline\urlprefix\url{http://
  www.slac.stanford.edu/econf/C010630/proceedings.shtm}

\bibitem{jlab2011}
J.~Grames, R.~Suleiman, P.~A. Adderley, J.~Clark, J.~Hansknecht, D.~Machie,
  M.~Poelker, M.~L. Stutzman, Charge and fluence lifetime measurements of a dc
  high voltage gaas photogun at high average current, Phys. Rev. ST Accel.
  Beams 14 (2011) 043501.

\bibitem{aulenbacher2011polarized}
K.~Aulenbacher, Polarized beams for electron accelerators, The European
  Physical Journal Special Topics 198~(1) (2011) 361.

\bibitem{aulenbacher1997mami}
K.~Aulenbacher, C.~Nachtigall, H.~Andresen, J.~Bermuth, T.~Dombo, P.~Drescher,
  H.~Euteneuer, H.~Fischer, D.~Harrach, P.~Hartmann, et~al., The mami source of
  polarized electrons, Nuclear Instruments and Methods in Physics Research
  Section A: Accelerators, Spectrometers, Detectors and Associated Equipment
  391~(3) (1997) 498--506.

\bibitem{grames2008biased}
J.~Grames, P.~Adderley, J.~Brittian, J.~Clark, J.~Hansknecht, D.~Machie,
  M.~Poelker, E.~Pozdeyev, M.~Stutzman, K.~Surles-Law, A biased anode to
  suppress ion back-bombardment in a dc high voltage photoelectron gun, in: AIP
  Conference Proceedings, Vol. 980, AIP, 2008, pp. 110--117.

\bibitem{grames2005ion}
J.~Grames, P.~Adderley, J.~Brittian, D.~Charles, J.~Clark, J.~Hansknecht,
  M.~Poelker, M.~Stutzman, K.~Surles-Law, Ion back-bombardment of gaas
  photocathodes inside dc high voltage electron guns, in: Particle Accelerator
  Conference, 2005. PAC 2005. Proceedings of the, IEEE, 2005, pp. 2875--2877.

\bibitem{clendenin2003slac}
J.~Clendenin, A.~Brachmann, T.~Galetto, D.-A. Luh, T.~Maruyama, J.~Sodja,
  J.~Turner, The slac polarized electron source, in: AIP Conference
  Proceedings, Vol. 675, AIP, 2003, pp. 1042--1046.

\bibitem{adderley2011cebaf}
P.~Adderley, J.~Clark, J.~Grames, J.~Hansknecht, M.~Poelker, M.~Stutzman,
  R.~Suleiman, K.~Surles-Law, J.~McCarter, M.~BastaniNejad, Cebaf 200kv
  inverted electron gun.

\bibitem{ptitsyn2017erl}
V.~Ptitsyn, et~al., Erl-ring and ring-ring designs for the erhic electron-ion
  collider, in: North American Particle Accelerator Conf.(NAPAC'16), Chicago,
  IL, USA, October 9-14, 2016, JACOW, Geneva, Switzerland, 2017, pp. 64--68.

\bibitem{bruening2013large}
O.~Bruening, M.~Klein, The large hadron electron collider, Modern Physics
  Letters A 28~(16) (2013) 1330011.

\bibitem{ptitsyn2016erl}
V.~Ptitsyn, I.~Pinayev, J.~S. Berg, S.~Belomestnykh, B.~Parker, R.~Than,
  M.~Blaskiewicz, K.~Brown, S.~Brooks, W.~Meng, et~al., The erl-based design of
  electron-hadron collider erhic.

\bibitem{abbott2016production}
D.~Abbott, P.~Adderley, A.~Adeyemi, P.~Aguilera, M.~Ali, H.~Areti, M.~Baylac,
  J.~Benesch, G.~Bosson, B.~Cade, et~al., Production of highly polarized
  positrons using polarized electrons at mev energies, Physical review letters
  116~(21) (2016) 214801.

\bibitem{tsentalovich2014status}
E.~Tsentalovich, Status of high intensity polarized electron gun project, PoS
  (2014) 043.

\bibitem{jlab2010}
P.~A. Adderley, J.~Clark, J.~Grames, J.~Hansknecht, K.~Surles-Law, D.~Machie,
  M.~Poelker, M.~L. Stutzman, R.~Suleiman, Load-locked dc high voltage gaas
  photogun with an inverted-geometry ceramic insulator, Phys. Rev. ST Accel.
  Beams 13 (2010) 010101.

\bibitem{dunham2013record}
B.~Dunham, J.~Barley, A.~Bartnik, I.~Bazarov, L.~Cultrera, J.~Dobbins,
  G.~Hoffstaetter, B.~Johnson, R.~Kaplan, S.~Karkare, et~al., Record
  high-average current from a high-brightness photoinjector, Applied Physics
  Letters 102~(3) (2013) 034105.

\bibitem{wang2018high}
E.~Wang, High current polarized electron source for future erhic, in: AIP
  Conference Proceedings, Vol. 1970, AIP Publishing, 2018, p. 050008.

\bibitem{Opera}
V.~F. Software, \href{https://operafea.com/product/}{Opera simulation
  software}.
\newline\urlprefix\url{https://operafea.com/product/}

\bibitem{liu2016effects}
W.~Liu, S.~Zhang, M.~Stutzman, M.~Poelker, Effects of ion bombardment on bulk
  gaas photocathodes with different surface-cleavage planes, Physical Review
  Accelerators and Beams 19~(10) (2016) 103402.

\bibitem{shutthanandan2012surface}
V.~Shutthanandan, Z.~Zhu, M.~L. Stutzman, F.~Hannon, C.~Hernandez-Garcia, M.~I.
  Nandasiri, S.~V. Kuchibhatla, S.~Thevuthasan, W.~P. Hess, Surface science
  analysis of gaas photocathodes following sustained electron beam delivery,
  Physical Review Special Topics-Accelerators and Beams 15~(6) (2012) 063501.

\bibitem{reiser2008}
M.~Reiser, Theory and design of charged particle beams, John Wiley \& Sons,
  2008.

\bibitem{gpt05}
S.~Van Der~Geer, O.~Luiten, M.~De~Loos, G.~P{\"o}plau, U.~Van~Rienen, 3d
  space-charge model for gpt simulations of high brightness electron bunches,
  in: Institute of Physics Conference Series, Vol. 175, 2005, p. 101.

\bibitem{ibb}
\href{https://github.com/biswas101/new-dynamic-ionizer-element}{Ionizer element
  for gpt}.
\newline\urlprefix\url{https://github.com/biswas101/new-dynamic-ionizer-element}

\bibitem{ion_trap}
E.~Pozdeyev, Ion trapping and cathode bombardment by trapped ions in dc
  photoguns, Phys. Rev. ST Accel. Beams 10 (2007) 083501.

\bibitem{bazarov2011comparison}
I.~V. Bazarov, A.~Kim, M.~N. Lakshmanan, J.~M. Maxson, Comparison of dc and
  superconducting rf photoemission guns for high brightness high average
  current beam production, Physical Review Special Topics-Accelerators and
  Beams 14~(7) (2011) 072001.

\end{thebibliography}

\end{document}